\documentclass[journal,10pt]{IEEEtran}
\usepackage{amsmath}
\usepackage{amssymb}
\usepackage{graphicx}
\usepackage{color}
\usepackage{cite}
\usepackage{bm}
\usepackage{epsfig,psfrag}
\usepackage{tabularx}
\usepackage{acronym}
\usepackage[yyyymmdd,hhmmss]{datetime}
\usepackage{bbm}
\usepackage{mathrsfs} 
\usepackage{tabularx}
\usepackage{multirow}
\usepackage{colortbl,pgfplotstable}
\usepackage{subfig}
\usepackage{tabularx}
\usepackage{multirow}
\usepackage{acronym}
\usepackage{tikz}

\acrodef{spa}[SPA]{sum-product algorithm}
\acrodef{da}[DA]{data association}
\acrodef{mmse}[MMSE]{minimum mean-square error}
\acrodef{pt}[PT]{potential target}
\acrodef{pmf}[pmf]{probability mass function}
\acrodef{pdf}[pdf]{probability density function}
\acrodef{iid}[iid]{independent and identically distributed}
\acrodef{rmse}[RMSE]{root-mean-squared error}
\acrodef{ospa}[OSPA]{optimal sub-pattern assignment}
\acrodef{bp}[BP]{belief propagation}
\acrodef{bpf}[BPF]{bootstrap particle filter}
\acrodef{upf}[UPF]{unscented particle filter}
\acrodef{pde}[PDE]{partial differential equation}
\acrodef{sde}[SDE]{stochastic differential equation}
\acrodef{ode}[ODE]{ordinary differential equation}
\acrodef{edh}[EDH]{exact Daum and Huang}
\acrodef{ledh}[LEDH]{localized exact Duam and Huang}
\acrodef{pfpf}[PFPF]{particle flow particle filter}
\acrodef{mcmc}[MCMC]{Markov Chain Monte Carlo}
\acrodef{smc}[SMC]{sequential Monte Carlo}
\acrodef{map}[MAP]{maximum a posteriori}
\acrodef{tdoa}[TDOA]{Time-difference of arrival}
\acrodef{pf}[PF]{Particle flow}
\acrodef{pda}[PDA]{probabilistic data association}
\acrodef{jpda}[JPDA]{Joint \ac{pda}}
\acrodef{phd}[PHD]{probability hypothesis density}
\acrodef{cphd}[CPHD]{cardinalized \ac{phd}}
\acrodef{mht}[MHT]{multi-hypothesis tracking}
\acrodef{slam}[SLAM]{simultaneous localization and mapping}
\acrodef{iid}[iid]{independent and identically distributed}
\acrodef{rfs}[RFS]{random finite sets}
\acrodef{gospa}[GOSPA]{generalized optimal subpattern assignment}

\newcommand{\rv}[1]{\mathsf{#1}}  
\newcommand{\V}[1]{\bm{#1}}
\newcommand{\RV}[1]{\boldsymbol{\mathsf{#1}}}

\newcommand{\M}[1]{\boldsymbol{\uppercase{#1}}}

\newcommand{\Set}[1]{\mathcal{#1}}

\providecommand{\rd}{\textcolor{black}}

\newcolumntype{L}[1]{>{\raggedright\arraybackslash}p{#1}}
\newcolumntype{C}[1]{>{\centering\arraybackslash}p{#1}}
\newcolumntype{R}[1]{>{\raggedleft\arraybackslash}p{#1}}

\providecommand{\be}{\begin{equation}}
\providecommand{\ee}{\end{equation}}

\providecommand{\ist}{\hspace*{.3mm}}

\providecommand{\rmv}{\hspace*{-.3mm}}

\providecommand{\nn}{\nonumber}
\newcommand{\T}{\text{T}}
\DeclareMathOperator*{\argmax}{argmax}

\allowdisplaybreaks
\definecolor{temporalgreen}{RGB}{0,128,0}
\definecolor{spatialred}{RGB}{255,0,0}
\definecolor{temporalblue}{RGB}{0,0,205}

\allowdisplaybreaks

\begin{document}
\title{\rd{Track Coalescence and Repulsion in Multitarget Tracking: An Analysis of MHT, JPDA, and \\Belief Propagation Methods}
\vspace*{2mm}}

\author{Thomas Kropfreiter,~\IEEEmembership{Member,~IEEE}, 
Florian Meyer,~\IEEEmembership{Member,~IEEE}, 
David F.~Crouse,~\IEEEmembership{Senior Member,~IEEE}, 
Stefano Coraluppi,~\IEEEmembership{Fellow,~IEEE}, 
Franz Hlawatsch,~\IEEEmembership{Fellow,~IEEE}, and 
Peter Willett,~\IEEEmembership{Fellow,~IEEE}  
\thanks{
This manuscript is dedicated to the memory of Dr.\ Craig A. Carthel, a brilliant applied mathematician and software architect who passed away
in July 2022. Dr.\ Carthel contributed to this work while employed with Systems \& Technology Research, Woburn, MA, USA.} 
\thanks{This work was supported in part by the University of California San Diego, in part by the Office of Naval Research under Grant N00014-23-1-2284, and in part by the Austrian Science Fund under Grant P32055-N31. This work was partly presented at FUSION-2021, Sun City, South Africa, Nov. 2021.}
\thanks{T.~Kropfreiter and F.~Meyer are with the Scripps Institution of Oceanography and the Department of Electrical and Computer Engineering, University of California San Diego, La Jolla, CA, USA (e-mail:\{tkropfreiter,\,flmeyer\} @ucsd.edu).}

\thanks{D. F.~Crouse is with the Naval Research Laboratory, Washington D.C., USA (e-mail: david.crouse@nrl.navy.mil).}

\thanks{S.~Coraluppi is with Systems \& Technology Research, Woburn, MA, USA (email: stefano.coraluppi@str.us).}

\thanks{F.~Hlawatsch is with the Institute of Telecommunications, TU Wien, Vienna, Austria (e-mail: franz.hlawatsch@tuwien.ac.at).}

\thanks{P.~Willett is with the Department of Electrical and Computer Engineering, University of Connecticut, Storrs, CT, USA (email: peter.willett@uconn.edu).}

\vspace*{-3mm}
}

\maketitle

\begin{abstract}
Joint probabilistic data association (JPDA) filter methods and multiple hypothesis tracking (MHT) methods are widely used  for multitarget tracking (MTT).  However, they are known to exhibit undesirable behavior in tracking scenarios with targets in close proximity: JPDA filter methods suffer from the track coalescence effect, i.e., the estimated tracks of targets in close proximity tend to merge and can become indistinguishable, while MHT methods suffer from an opposite effect known as track repulsion\rd{, i.e., the estimated tracks of targets in close proximity tend to repel each other in the sense that their separation is larger than the actual distance between the targets.} In this paper, we review the JPDA filter and MHT methods and discuss the track coalescence and track repulsion effects. We also consider a more recent methodology for MTT that is based on the belief propagation (BP) algorithm. 
\rd{We argue that BP-based MTT does not exhibit track repulsion because it is not based on maximum a posteriori estimation, and that it exhibits significantly reduced track coalescence because certain properties of the BP messages related to data association encourage separation of target state estimates.}
Our theoretical arguments are confirmed by numerical results \rd{for four representative simulation scenarios.}
\vspace{0mm}\end{abstract}

\begin{IEEEkeywords}
Multitarget tracking, joint probabilistic data association, JPDA, multiple hypothesis tracking, MHT, belief propagation,
track coalescence, track repulsion.
\end{IEEEkeywords}

\acresetall
\section{Introduction}\label{sec:intro}

\subsection{Background and Motivation}
\label{sec:intro_background}

The goal of multitarget tracking (MTT) is to recursively estimate the time-dependent number and states of multiple objects (``targets'') based on noisy sensor measurements \cite{barShalom11,reid79,ChaMor11,mahler2007statistical,koch14,mey18}. This enables situational awareness in a wide variety of applications, including aerospace and maritime surveillance \cite{barShalom74,reid79,bar-shalom09,fortmann83,FerMunTesBraMeyPelPetAlvStrLeP:J17,Kim21,Gag20} and robotics \cite{levinson11,urmson08,MeyWil:J21}. However, MTT is a challenging task due to clutter, missed detections, and measurement-origin uncertainty \cite{barShalom11,ChaMor11,mahler2007statistical,koch14,mey18}. MTT methods are typically placed in the framework of Bayesian inference and can be broadly classified as vector-type and set-type methods. Vector-type methods \cite{barShalom11,barShalom74,Bar:B90,reid79,bar-shalom09,fortmann83,musicki04,VerMas05,MusLaS08,horridge06,PatPop00,Cox96,DanNew06,Bla04,Mor77,deb97,ZhaMey:J24,LiaMey:J24} model the multitarget state by a random vector, whereas set-type methods \cite{mahler2007statistical,Mah03,vo05,vo07,williams2015marg,Nannuru16,Gar20SoT,Gra18SoTPMBM,KroMeyHla:J20,KroMeyHla:J22} model it by a random finite set.

Two classical and popular instances of vector-type MTT methods are the joint probabilistic data association (JPDA) filter \cite{barShalom74,fortmann83,barShalom11} and multiple hypothesis tracking (MHT) methods \cite{reid79,PatPop00,Cox96,DanNew06,Bar:B90,Bla04,Mor77}. The JPDA filter is a single-scan method that seeks to calculate minimum mean square error (MMSE) state estimates for a known number of targets at each time step. An unknown number of targets can be addressed using heuristics \cite{barShalom11} or a probabilistic model for target existence as used for the development of the joint integrated probabilistic data association (JIPDA) filter \cite{musicki04}. \Ac{da} hypotheses are incorporated in a soft (probabilistic) manner by calculating weighted sums over all possible associations at each time step. This is equivalent to ``marginalizing out'' the unknown \ac{da} vector in the posterior \ac{pdf} of the target states, which can also be seen as an exhaustive averaging over \ac{da} vectors. In contrast, MHT methods are multiscan methods that achieve hard \ac{da} by performing approximate maximum a posteriori (MAP) detection of an entire sequence of \ac{da} vectors using optimization techniques such as the Auction algorithm or multiple frame assignment \cite{Ber:B91}. Subsequently, the target states are estimated individually, e.g., using Kalman smoothing. Both MHT and JPDA methods are capable of maintaining track continuity, i.e., target identification over consecutive time steps. Besides the aspect of performing exhaustive averaging as in JPDA filter methods or approximate maximization as in MHT methods, an additional aspect is the representation of a single potential target or ``track''. Typical representations are by a single Gaussian \cite{barShalom74,reid79}, a Gaussian mixture \cite{Sal:J90,Pao:94}, or a set of particles \cite{vermaak05,mey17}. We note that certain set-type filters, e.g.,  \cite{williams2015marg} and \cite{KroMeyHla:J22}, also use exhaustive averaging or approximate maximization as their DA strategy.

Experimental evidence has demonstrated undesirable behavior of JPDA and MHT type methods in tracking scenarios with targets in close proximity \cite{fitzgerald85,corcar09}. More specifically, JPDA filter methods suffer from the track coalescence effect \cite{fitzgerald85}, which means that the estimated tracks of closely spaced targets that move in parallel tend to come together and merge, thereby becoming indistinguishable. In MHT methods, on the other hand, an opposite effect called track repulsion can be observed \cite{corcar09}: the estimated tracks of closely spaced targets that move in parallel repel each other in the sense that their separation is larger than the actual distance between the targets. Methodological modifications intended to mitigate these detrimental effects were proposed in \cite{bloblo00,bloblo15,corcar12,SveSve11,Wil14,LauWil16}. In particular, track coalescence can be mitigated by using a specific hypothesis pruning strategy \cite{bloblo00,bloblo15} or filter designs based on the \ac{ospa} metric \cite{SveSve11,Wil14} or on variational methods \cite{LauWil16}, and track repulsion by choosing a \ac{da} hypothesis from a MAP equivalence class rather than the single MAP solution \cite{corcar12}. However, these modifications are unsuitable for large-scale tracking scenarios with an unknown and time-varying number of targets.

Among the set-type MTT methods, some avoid \ac{da} at the cost of using a suboptimal postprocessing step for track formation and, possibly, exhibiting a reduced target detection and tracking performance \cite{Mah03,vo05,vo07,Nannuru16}. Others rely on \ac{da} strategies similar to those used by the JPDA filter and hence also suffer from track coalescence \cite{reuter14,williams2015marg}. Inspired by the concept of set-type MTT, track coalescence has also been investigated in the context of permutation-invariant estimation \cite{CroWilBar:C11,CroWilGue:C11,SveSve11}.

A recently developed methodology for vector-type MTT describes filtering and \ac{da} by a factor graph that provides a blueprint for applying belief propagation (BP) \cite{williams14,mey17,mey18,MeyWil:J21,SolMeyBraHla:J19,Gag20}. BP, also known as the \textit{sum-product algorithm} \cite{kschischang01,koller09}, is based on a factor graph \cite{kschischang01,koller09} and aims at computing the marginal posterior pdfs \rd{or \acp{pmf}} in an efficient way. For each node of the factor graph, certain messages are calculated, each of which is passed to some of the adjacent nodes. If the factor graph is a tree, i.e., without loops, then the solutions provided by BP---referred to as beliefs---are equal to the marginal posterior pdfs \rd{or \acp{pmf}}. For factor graphs with loops, BP has to be performed iteratively, and the beliefs are only approximations of the respective marginal posterior pdfs \rd{or \acp{pmf}}; these approximations have been observed to be very accurate in many applications \cite{kschischang01,koller09}. BP can be applied for the inference of discrete and continuous random variables. If the belief represents a marginal pmf, its values can be interpreted as probabilities. In the context of probabilistic DA in MTT, BP is guaranteed to converge \cite{williams14}.

By performing efficient approximate marginalization operations within the overall MMSE estimation framework of the JPDA filter, the BP algorithm provides scalable solutions to both probabilistic \ac{da}---accounting for measurement-origin uncertainty, clutter, and missed detections---and nonlinear state estimation for randomly appearing and disappearing targets. BP-based MTT methods exhibit excellent scalability in the number of targets and in the number of measurements because the BP algorithm systematically exploits conditional independence properties of the underlying statistical model for a reduction of computational complexity. BP-based methods are also capable of maintaining track continuity. Since the nodes of the graph corresponding to the target states are ordered, BP-based methods are quite different from the set-type approach that aims at performing permutation-invariant or unordered estimation.  BP methods rely on efficient approximations to exhaustive averaging and are thus related to J(I)PDA filters. However, contrary to J(I)PDA filters, BP methods employ an improved birth model that, together with the efficient DA solution, allows the detection and tracking of a potentially very large number of targets even if the targets are in close vicinity in space and time.

\vspace{-1mm}
\subsection{Contributions and Paper Organization} 
\label{sec:intro_contr}

In this paper, we review the track coalescence and track repulsion effects and investigate them in the context of the JPDA filter and MHT methods as well as the BP-based MTT methods. Our analysis suggests that BP-based MTT methods do not suffer from the track repulsion effect, while the track coalescence effect is significantly smaller than in JPDA filter methods. These findings are confirmed by simulation results for four different MTT scenarios, which demonstrate a significant reduction of track coalescence and an absence of track repulsion in BP-based methods relative to JPDA filter methods and MHT methods, respectively. In contrast to permutation-invariant estimation \cite{CroWilBar:C11,CroWilGue:C11,SveSve11}, where track coalesence is avoided by an alternative formulation of the estimation problem, in BP-based MTT, track coalescence is reduced because certain properties of the DA-related BP messages encourage separation of target state estimates.

The main contributions of the paper are as  
\vspace{.5mm}
follows:
\begin{itemize}
\item We discuss and analyze the track coalescence effect and the track repulsion effect in the context of the JPDA filter and MHT methods, respectively.
\vspace{.8mm}
\item We demonstrate numerically that BP-based MTT does not suffer from the track repulsion effect, while the track coalescence effect is much smaller than in the JPDA filter.\vspace{.8mm} 
\item We provide evidence that the reduced coalescence in BP-based MTT is due to the nature of the DA solution and not due to the choice of track representation.
\vspace{.5mm}
\end{itemize}

The remainder of this paper is organized as follows. 
The basic notation is described in the next subsection. 
In Section \ref{sec:systemModel}, we present the MTT system model used in this paper as well as a statistical formulation.
\rd{Section \ref{sec:JointAssociation} describes the resulting joint posterior distribution and the corresponding factor graph.}
The track repulsion effect in MHT methods and the track coalescence effect in JPDA filter methods are considered in 
Sections \ref{sec:MHT-Tracking} and \ref{sec:JPDA-Tracking}, respectively. In Section~\ref{sec:BP-Tracking}, we describe the BP approach to MTT.
The reduction of the track coalescence effect achieved by particle-based BP methods is analyzed in Section \ref{sec:closerlook}.
Finally, simulation results are presented in Section \ref{sec:results}.
We note that this paper advances beyond our earlier conference publication \cite{Kro21BP} in that it contains an improved introduction to MHT and JPDA filter methods; it presents a simple illustration of the track repulsion effect due to MHT-like hard \ac{da}; it provides a detailed analysis of the track coalescence effect and an explanation why BP methods are less susceptible to it; and it presents a more extensive simulation analysis.
    
\vspace{-1mm}

\subsection{Notation} 
\label{sec:intro_notation}

Random variables are displayed by sans serif, upright print (e.g., $\rv x$) and their realizations by serif, italic print (e.g., $x$). 
Vectors are denoted by bold lowercase letters (e.g., $\RV x$ or $\V x$) and matrices by bold uppercase letters (e.g., $\M{A}$).
Furthermore, ${\V{x}}^\T$ denotes the transpose of vector $\V x$, $\propto$ indicates equality up to a normalization factor, and $f(\V x)$ denotes the \ac{pdf} and $p(\V x)$ the \ac{pmf} of the, respectively, continuous and discrete random vector $\RV{x}$. 
We also use $f(\V{x})$ to denote the hybrid pdf/pmf of a mixed continuous-discrete random vector $\RV{x}$. Finally, $1(a)$ is defined to be $1$ if $a \rmv=\rmv 0$ and $0$ otherwise, and $\delta({x})$ denotes the Dirac delta function.

\section{System Model}
\label{sec:systemModel}

We consider the MTT system model from \cite{barShalom11,ChaMor11,mey18,Bla04}. This model and the corresponding statistical formulation will be briefly described for completeness.

\vspace{-1mm}
\subsection{Potential Target States and State Evolution}
\label{sec:systemModel_states}

The number of targets is time-varying and unknown, which is accounted for by using the concept of \acp{pt} \cite{mey17,mey18}. At discrete time $k \rmv=\rmv 0,1,\ldots$, the number of \acp{pt} $J_k$ is the maximum possible number of targets, which is determined as described in Section \ref{sec:DA-MLF}. The state $\big[\RV{x}^{(j)\T}_{k}\; \rv{r}^{(j)}_{k}\big]^\T\rmv$ of \ac{pt} $j \rmv\in\rmv \{1,\dots,J_k\}$ at time $k$ consists of a kinematic state $\RV{x}^{(j)}_{k}\!\rmv\in\rmv \mathbb{R}^{n}\rmv$, which comprises the \ac{pt}'s position and possibly further motion-related parameters, and a binary existence indicator variable $\rv{r}^{(j)}_{k} \!\in\rmv \{0,1\}$ that is $1$ if \ac{pt} $j$ exists at time $k$ and $0$ otherwise.  This existence indicator variable was first introduced in the context of the JIPDA filter \cite{musicki04}.  The kinematic state $\RV{x}^{(j)}_{k}$ of a nonexisting \ac{pt}---i.e., of a \ac{pt} with $\rv{r}^{(j)}_{k} \!\rmv=\rmv 0$---is undefined; its distribution will therefore 
be described by an arbitrary ``dummy \ac{pdf}\ist'' $f_{\text{D}}\big(\V{x}^{(j)}_{k}\big)$.\linebreak 
Accordingly, all \acp{pdf} of \ac{pt} states, $ f\big(\V{x}^{(j)}_{k}\rmv, r^{(j)}_{k}\big)$, have the property that $f\big(\V{x}^{(j)}_{k}\rmv, r^{(j)}_{k} \!\!=\!0 \big) \rmv=\rmv f^{(j)}_{k} f_{\text{D}}\big(\V{x}^{(j)}_{k}\big)$, where $f^{(j)}_{k} \!\rmv\in\rmv [0,1]$ is the probability of nonexistence.  \rd{Only targets that exist can generate measurements.}

For each \ac{pt} state $\big[\RV{x}^{(j)\T}_{k-1}\; \rv{r}^{(j)}_{k-1}\big]^\T\rmv$, $j \rmv\in\rmv \{1,\dots,J_{k-1}\}$ at time $k\rmv-\rmv1$, there is one ``legacy'' \ac{pt} state $\big[\underline{\RV{x}}^{(j)\T}_{k}\; \underline{\rv{r}}^{(j)}_{k}\big]^\T\rmv$ at time $k$. The motion and potential disappearance of \ac{pt} $j$ are modeled statistically by the single-target state-transition \ac{pdf}/\ac{pmf} $f\big( \underline{\V{x}}_k^{(j)}\rmv\rmv, \underline{r}_k^{(j)} \big| \V{x}_{k-1}^{(j)}, r_{k-1}^{(j)} \big)$, which involves the kinematic state-transition \ac{pdf} $f\big( \underline{\V{x}}_k^{(j)} \big| \V{x}_{k-1}^{(j)} \big)$ and the probability of target survival $p_{\text{s}}$ as described in, e.g., \cite[Section VIII-C]{mey18}. All \ac{pt} states evolve independently, and they are independent at time $k \rmv=\rmv 0$. In the absence of any information about the number of PTs at time $k \rmv=\rmv 0$, we set $J_0 \rmv=\rmv 0$.

\vspace{-1mm}
\subsection{Measurements, New \acp{pt}, and Data Association}
\label{sec:DA-MLF}

At each time (or ``scan'') $k\rmv\geq\rmv1$, a sensor produces $\rv{M}_k$ measurements $\RV{z}^{(m)}_{k}\rmv$, $m \rmv\in\rmv \big\{ 1,\dots,\rv{M}_k \big\}$. Here, $\rv{M}_k$ is modeled as a random variable whose value (realization) $M_k$ is known once the measurements have been observed. We define the joint vector of all measurements at time $k$ as $\RV{z}_k \triangleq \big[ \RV{z}^{(1)\T}_k \rmv\cdots\ist \RV{z}^{(\rv{M}_k)\T}_k\big]^{\mathrm{T}}\!$. Each measurement $\RV{z}^{(m)}_{k}$ has one of three possible origins:
(i) a new \ac{pt} representing a target that generates a measurement for the first time, or 
(ii) a legacy \ac{pt} representing a target that generated at least one measurement before, or 
(iii) clutter.

The birth of new targets is modeled by a Poisson point process with mean parameter $\mu_{\text{b}}$ (which is the mean number of newborn targets) and spatial \ac{pdf} $f_{\text{b}}\big( \overline{\V{x}}_{k} \big)$. To account for the birth of new targets, at time time $k$, $\rv{M}_k$ new \ac{pt} states $\big[\overline{\RV{x}}^{(m)\T}_{k}\; \overline{\rv{r}}^{(m)}_{k}\big]^\T\rmv$, $m \rmv\in\rmv \{1,\dots,\rv{M}_k\}$ are introduced, where $\overline{\rv{r}}^{(m)}_{k} \!=\! 1$ means that measurement $\RV{z}^{(m)}_{k}$ has origin (i) and $\overline{\rv{r}}^{(m)}_{k} \!=\! 0$ means that it has origin (ii) or (iii).

We define the joint vector of all kinematic states at time $k$ as $\RV{x}_k \rmv\triangleq\rmv \big[\RV{x}_k^{(1)}\rmv\cdots\ist\RV{x}_k^{(J_k)}\big]^{\T} \rmv\rmv=\rmv  \big[\underline{\RV{x}}^{(1)\T}_k \rmv\cdots\ist \underline{\RV{x}}^{(J_{k-1})\T}_k \ist\ist\ist \overline{\RV{x}}^{(1)\T}_k \rmv\cdots\ist \overline{\RV{x}}^{(\rv{M}_k)\T}_k\big]^{\mathrm{T}}\!$, and the joint vector of all existence indicators at time $k$ as $\RV{r}_k \triangleq \big[\rv{r}_k^{(1)}\rmv\cdots\ist \rv{r}_k^{(J_k)} \big]^{\T} \rmv\rmv=\rmv \big[\underline{\rv{r}}^{(1)}_k \rmv\cdots\ist \underline{\rv{r}}^{(J_{k-1})}_k \ist\, \overline{\rv{r}}^{(1)}_k \rmv\cdots\ist \overline{\rv{r}}^{(\rv{M}_k)}_k\big]^{\mathrm{T}}\!$. It follows that once the measurements have been observed and, thus, $M_{k}$ is known, the total number of \acp{pt}---both legacy \acp{pt} and new \acp{pt}---is $J_{k} \rmv=\rmv J_{k-1} \rmv+\rmv M_{k}$. 

The target represented by \ac{pt} $j$ is detected by the sensor, in the sense that it generates a measurement $\RV{z}_k^{(m)}\rmv$, with probability $p_{\text{d}}$. The dependence of a target-generated measurement $\RV{z}_k^{(m)}$ on the kinematic state $\RV{x}_k^{(j)}$ of the corresponding detected \ac{pt} is modeled statistically by the conditional \ac{pdf} $f\big(\V{z}^{(m)}_{k} | \V{x}^{(j)}_{k} \big)$, which can be derived from the measurement model of the sensor. Clutter measurements are modeled by a Poisson point process with mean parameter $\mu_{\text{c}}$ (which is the mean number of clutter measurements) and spatial \ac{pdf} $f_{\text{c}}\big( \V{z}^{(m)}_{k} \big)$. Note that if $f_{\text{c}}\big( \V{z}^{(m)}_{k} \big)$ is uniform, then the Poisson point process has a constant spatial density.

The origin of each measurement is unknown, i.e., it is not known if the measurement is generated by a \ac{pt}, and by which \ac{pt}, or if it is clutter. We use the conventional \ac{da} assumption, which postulates that a target can generate at most one measurement, and a measurement can be generated by at most one target \cite{barShalom11,mahler2007statistical,mey18}. The association between the $M_k$ measurements and the $J_{k-1}$ legacy \acp{pt} can be modeled by the ``target-oriented'' \ac{da} vector $\RV{a}_{k} = \big[\rv{a}_{k}^{(1)} \rmv\cdots\ist \rv{a}_{k}^{(J_{k-1})} \big]^{\T}\rmv$ whose $j$th entry $\rv{a}_{k}^{(j)}$ is $m \in \{1,\dots,M_{k} \}$ if legacy \ac{pt} $j$ generated measurement $m$ and zero if it did not generate any measurement \cite{barShalom11,mey18}. \rd{Since only targets that exist can generate measurements, $a_{k}^{(j)} \!\in \{1,\dots,M_{k} \}$ implies that $\underline{r}_{k}^{(j)} = 1$. However, since there can be a missed detection, we can have $a_{k}^{(j)}  = 0$ for  $\underline{r}_{k}^{(j)} = 1$. Note that $a_{k}^{(j)}  = m \in \{1,\ldots,M_k\}$ implies that measurement $\RV{z}^{(m)}_{k}$ has origin (ii).} A DA vector $\V{a}_{k}$ is considered ``valid'' if it satisfies the DA assumption mentioned above. \rd{Due to the DA assumption, a DA vector $\V{a}_{k}$ is valid if and only if at most one entry is equal to any specific $m \in \{1,\dots,M_{k} \}$. Also note that due to the DA assumption, $a^{(j)}_{k} = m$ implies $\overline{r}_{k}^{(m)} = 0$, because if measurement $m$ was originated by a legacy PT, the corresponding new PT cannot correspond to an actual target.}

We also introduce the ``measurement-oriented'' \ac{da} vector $\RV{b}_{k} = \big[\rv{b}_{k}^{(1)} \rmv\cdots\ist \rv{b}_{k}^{(M_k)} \big]^{\T}\rmv$ whose $m$th entry $\rv{b}_{k}^{(m)}$ is $j \in \{1,\dots,J_{k-1} \}$ if measurement $m$ was generated by legacy \ac{pt} $j$ and zero if it was generated by a new \ac{pt} or by clutter. Together, $\RV{a}_{k}$ and $\RV{b}_{k}$ constitute a redundant DA representation because for any given valid $\V{a}_{k}$, there is exactly one valid $\V{b}_{k}$ and for any given valid $\V{b}_{k}$, there is exactly one valid $\V{a}_{k}$. However, this redundant representation makes it possible to develop scalable MTT methods that exploit the structure of the DA problem for a substantial reduction of computational complexity \cite{mey18,williams14} (see Section \ref{sec:BP-Tracking}).

\vspace{-1mm}
\section{\rd{Joint Posterior Distribution and\\Factor Graph}}
\label{sec:JointAssociation}

Let $\V{x}_{0:k}$ denote the vector obtained by stacking the vectors $\V{x}_{0}$ through $\V{x}_{k}$, and similarly for $\V{r}_{0:k}$, $\V{a}_{1:k}$, $\V{b}_{1:k}$, and $\V{z}_{1:k}$ as well as their random counterparts. Using common assumptions, it has been shown in \cite{williams2015marg,williams14,mey17,mey18} that the joint posterior \ac{pdf}/\ac{pmf} of $\RV{x}_{0:k}$, $\RV{r}_{0:k}$, $\RV{a}_{1:k}$, and $\RV{b}_{1:k}$ conditioned on the observed (and thus fixed) measurements $\V{z}_{1:k}$ is given by
\begin{align}
&f( \V{x}_{0:k}, \V{r}_{0:k}, \V{a}_{1:k}, \V{b}_{1:k} | \V{z}_{1:k} ) \nn\\[0mm]
&\hspace{-2mm}\propto \rmv  \Bigg(\rmv \prod^{J_{0}}_{j=1} \! f\big(\V{x}^{(j)}_{0} \!, r^{(j)}_{0}  \big)\rmv  \rmv\bigg)\rmv \rmv \rmv  \prod^{k}_{k'=1} \!\rmv  g(\V{x}_{k'}, \V{r}_{k'},\V{a}_{k'}, \V{b}_{k'}, \V{x}_{k'\!-1}, \V{r}_{k'\!-1};  \V{z}_{k'\!}) , \nn \\[-3mm]
  \label{eq:jointPosteriorComplete} \\[-7.5mm]
\nn
\end{align}
with
\begin{align}
&g(\V{x}_{k}, \V{r}_{k}, \V{a}_{k}, \V{b}_{k}, \V{x}_{k-1}, \V{r}_{k-1};  \V{z}_{k\!}) \nn\\[.5mm]
&\hspace{2mm} \triangleq \Bigg(\prod^{J_{k\!-1}}_{j'=1} \! 
    f\big(\underline{\V{x}}^{(j')}_{k} \!,\underline{r}^{(j')}_{k} \big| \V{x}^{(j')}_{k-1},r^{(j')}_{k-1}\big) \rmv\Bigg) \nn\\[0.1mm] 
  &\hspace{7mm}\times \hspace{-.5mm}\Bigg( \prod^{J_{k\!-1}}_{j=1} \! q\big( \underline{\V{x}}^{(j)}_{k}\!, \underline{r}^{(j)}_{k}\!, a^{(j)}_{k}\rmv; \V{z}_{k} \big)\rmv\rmv 
    \prod^{M_{k}}_{m'=1} \!\! \Psi_{j,m'}\big(a_{k}^{(j)}\rmv, b_{k}^{(m')}\big) \rmv\Bigg)  \nn\\[.5mm]
  &\hspace{7mm}\times \hspace{-.5mm} \prod^{M_{k}}_{m=1} \! v\big( \overline{\V{x}}^{(m)}_{k}\!, \overline{r}^{(m)}_{k}\!, b^{(m)}_{k}\rmv; \V{z}_{k}^{(m)} \big)\ist . \label{eq:marginalization2_1} 
\end{align}
Here, the factor $q\big( \underline{\V{x}}^{(j)}_{k}\!, \underline{r}^{(j)}_{k}\!, a^{(j)}_{k} \rmv; \V{z}_{k} \big)$ is given by
\begin{align}
&\hspace{-3mm}q\big( \underline{\V{x}}^{(j)}_{k}\!, 1, a^{(j)}_{k}\rmv; \V{z}_{k} \big) \nn\\[1mm]
&\hspace{-2mm}= \begin{cases}
    \frac{ p_{\text{d}} }{ \mu_{\text{c}} f_{\text{c}}( \V{z}_{k}^{(m)} )} \ist f\big( \V{z}_{k}^{(m)} \rmv\big|\ist \underline{\V{x}}_{k}^{(j)} \big), & a^{(j)}_{k} \!=\rmv m \rmv\in\rmv \{1,\dots,M_k \} \\[2.5mm]
     1 \!-\! p_{\text{d}}  \ist, & a^{(j)}_{k} \!=\rmv 0 
  \end{cases} 
	\nn\\[-10mm]
	\label{eq:qFunction} \\[-1mm]
  \nn
\end{align}
and $q\big( \underline{\V{x}}^{(j)}_{k}\!, 0, a^{(j)}_{k}\rmv; \V{z}_{k} \big) \rmv=\rmv 1(a^{(j)}_{k})$, and the factor $v\big( \overline{\V{x}}^{(m)}_{k}\!, \overline{r}^{(m)}_{k}\!, \linebreak 
b^{(m)}_{k}\rmv; \V{z}_{k}^{(m)} \big)$ is given by
\begin{align}
&\hspace{-3mm}v\big( \overline{\V{x}}^{(m)}_{k}\!, 1, b^{(m)}_{k}\rmv; \V{z}_{k}^{(m)} \big) \nn \\[1mm]
&\hspace{-2mm}= \begin{cases}
     0 \ist,  & b^{(m)}_{k} \!\rmv\in\rmv \{ 1,\dots,J_{k-1} \}\\[1mm]
     \frac{p_{\text{d}} \ist \mu_{\text{b}} \ist f_{\text{b}}(\overline{\V{x}}^{(m)}_{k})}{ \mu_{\text{c}} \ist f_{\text{c}}( \V{z}_{k}^{(m)} )} \ist f\big(\V{z}_k^{(m)}  \big| \overline{\V{x}}^{(m)}_{k} \big) \ist, 
       & b^{(m)}_{k} \!=\rmv 0 
  \end{cases} 
		\nn\\[-10mm]
	\label{eq:vFunction} \\[-1mm]
  \nn
\end{align}
and $v\big( \overline{\V{x}}^{(m)}_{k}\!, 0, b^{(m)}_{k}\rmv; \V{z}_{k}^{(m)} \big) \rmv=\rmv f_{\text{D}}\big(\overline{\V{x}}^{(m)}_{k}\big)$. Furthermore, $\Psi_{j,m}\big(a_{k}^{(j)}\rmv,b_{k}^{(m)}\big)$ is a binary indicator function that checks the consistency of the target-oriented \ac{da} variable $a_{k}^{(j)}$ and the measurement-oriented \ac{da} variable $b_{k}^{(m)}$: it is zero if $a_{k}^{(j)} \!\rmv=\rmv m$, $b^{(m)}_{k} \!\rmv\neq\rmv j$ or $b^{(m)}_{k} \!\rmv=\rmv j$, $a_{k}^{(j)} \!\rmv\neq\rmv m$ and one otherwise (see \cite{williams14,mey18} for details). 
\rd{This consistency of the \ac{da} variables $a_{k}^{(j)}$ and $b_{k}^{(m)}$ guarantees that the DA assumption is not violated.}
The $\propto$ sign in \eqref{eq:jointPosteriorComplete} has to be interpreted in the sense that the right-hand side, when multiplied by a normalization constant that does not depend on $\V{x}_{0:k}$, $\V{r}_{0:k}$, $\V{a}_{1:k}$, or $\V{b}_{1:k}$, is equal to the joint posterior pdf/pmf $f( \V{x}_{0:k}, \V{r}_{0:k}, \V{a}_{1:k}, \V{b}_{1:k} | \V{z}_{1:k} )$. We note that the factor $g(\V{x}_{k}, \V{r}_{k}, \V{a}_{k}, \V{b}_{k}, \V{x}_{k-1}, \V{r}_{k-1};  \V{z}_{k})$ in \eqref{eq:jointPosteriorComplete} groups all factors corresponding to a single time step; this representation will support future discussions in Section \ref{sec:BP-Tracking}. A detailed derivation of expressions \eqref{eq:jointPosteriorComplete}--\eqref{eq:vFunction} is provided in \cite[Sec.~VIII-G]{mey18}. 
 
The factorization in \eqref{eq:jointPosteriorComplete} presupposes that the measurements $\V{z}_{1:k}$ are observed; accordingly, we use the semicolumn ``;''  in \eqref{eq:marginalization2_1}--\eqref{eq:vFunction} to indicate that the argument to the right of  ``;''  is fixed. The observed measurements $\V{z}_{1:k}$ reveal additional factorization structures unavailable in the statistical model with $\V{z}_{1:k}$ random. Since for Bayesian estimation we are either interested in marginal posterior \acp{pdf}/\acp{pmf} or values of $\V{x}_{0:k}$, $\V{r}_{0:k}$, $\V{a}_{1:k}$, and $\V{b}_{1:k}$ that maximize the joint posterior \ac{pdf}/\ac{pmf}, we do not need to compute the normalization constant. In case we want to obtain marginal posterior pdfs/pmfs, we can first marginalize the unnormalized version of the joint posterior pdf/pmf on the right-hand side of \eqref{eq:jointPosteriorComplete} and then normalize the result after marginalization. Since the marginalization step reduces the dimensionality, computing a normalization constant after marginalization always has a lower computational complexity.

\begin{figure}[t!]
\centering\hspace{4.5mm}
\vspace{-1mm}
\centering
\includegraphics[scale=.75]{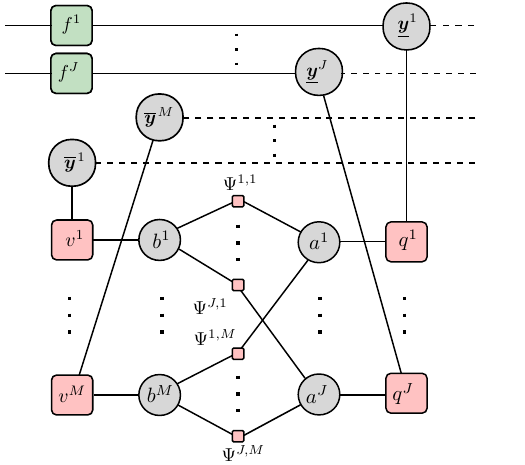}
\vspace{2mm}
 \caption{Factor graph representing the joint posterior pdf/pmf $f( \V{x}_{0:k},\V{r}_{0:k}$, $\V{a}_{1:k}, \V{b}_{1:k} | \V{z}_{1:k} )$ in \eqref{eq:jointPosteriorComplete}, \eqref{eq:marginalization2_1} for  one time step $k$. The time index $k$ is omitted, and the following short notations are used: 
$M \rmv\triangleq M_{k}$, 
$J \rmv\triangleq J_{k-1}$,
$\underline{\V{y}}^{j} \rmv\triangleq\rmv [\underline{\V{x}}^{(j)\T}_k\ist\ist \underline{r}^{(j)}_k]^{\T}$,
$\overline{\V{y}}^{m} \rmv\triangleq\rmv [\overline{\V{x}}^{(m)\T}_k \ist\ist\overline{r}^{(m)}_k]^{\T}$,
$a^j \rmv\triangleq a^{(j)}_k$,
$b^m \rmv\triangleq b^{(m)}_k$,
$f^j \rmv\triangleq f\big(\underline{\V{x}}^{(j)}_k,\underline{r}^{(j)}_k \big|\V{x}^{(j)}_{k-1},r^{(j)}_{k-1}\big)$,
$q^j \rmv\triangleq q\big( \underline{\V{x}}^{(j)}_{k}\rmv, \underline{r}^{(j)}_{k}\rmv, a_{k}^{(j)};\V{z}_{k} \big)$, 
$v^m \rmv\triangleq v\big( \overline{\V{x}}^{(m)}_{k} \rmv,\overline{r}^{(m)}_{k} \rmv,b_{k}^{(m)};\V{z}^{(m)}_{k} \big)$,
$\Psi^{j,m} \rmv\triangleq \Psi_{j,m}\big(a^{(j)}_{k} \rmv,b^{(m)}_{k}\big)$.}
  \label{fig:factorGraph}
\vspace{-2mm}
\end{figure}

The factorization of the joint posterior pdf/pmf $f( \V{x}_{0:k}, \V{r}_{0:k}, \linebreak 
\V{a}_{1:k}, \V{b}_{1:k} | \V{z}_{1:k} )$ as given by \eqref{eq:jointPosteriorComplete}, \eqref{eq:marginalization2_1} is represented graphically by the factor graph \cite{koller09,kschischang01} shown in Fig.~\ref{fig:factorGraph}. In this factor graph, the random variables and pdfs/pmfs involved in $f( \V{x}_{0:k}, \V{r}_{0:k}, \V{a}_{1:k}, \V{b}_{1:k} | \V{z}_{1:k} )$ are represented by circles and squares, respectively  \cite{kschischang01}. A circle is connected with a square if the random variable represented by the circle is involved in the pdf/pmf represented by the square \cite{kschischang01}.  Note that in the factor graph, we combine the kinematic state of a single target and the corresponding existence indicator variable into a single state variable  $\underline{\V{y}}^{j} \rmv\triangleq\rmv [\underline{\V{x}}^{(j)\T}_k\ist \underline{r}^{(j)}_k]^{\T}$ or $\overline{\V{y}}^{m} \rmv\triangleq\rmv [\overline{\V{x}}^{(m)\T}_k \ist\ist\overline{r}^{(m)}_k]^{\T}$, which is represented by a single variable node.

Marginalizing out the redundant measurement-oriented \ac{da} vector $\RV{b}_{1:k}$ from $f( \V{x}_{0:k}, \V{r}_{0:k}, \V{a}_{1:k}, \V{b}_{1:k} | \V{z}_{1:k} )$ yields the posterior pdf/pmf 
\begin{align}
f( \V{x}_{0:k}, \V{r}_{0:k}, \V{a}_{1:k} | \V{z}_{1:k} ) = \sum_{\V{b}_{1:k}} f( \V{x}_{0:k}, \V{r}_{0:k}, \V{a}_{1:k}, \V{b}_{1:k} | \V{z}_{1:k} ) \ist. \nn\\[-3.5mm]
 \label{eq:jointPosteriorComplete1} \\[-7mm]
\nn
\end{align}
This marginalization is in fact trivial since for a given $\V{a}_{1:k}$ there is exactly one $\V{b}_{1:k}$ for which \eqref{eq:jointPosteriorComplete} is nonzero. The posterior pdf/pmf $f( \V{x}_{0:k}, \V{r}_{0:k}, \V{a}_{1:k} | \V{z}_{1:k} )$ in \eqref{eq:jointPosteriorComplete1} together with expression \eqref{eq:jointPosteriorComplete}, \eqref{eq:marginalization2_1} represents a system model that is identical to the one used by MHT methods\footnote{In the MHT literature, the two vectors $\RV{r}_{0:k}$ and $\RV{a}_{1:k}$ are typically represented by a single equivalent vector $\RV{q}_{0:k}$ referred to as a global hypothesis.} and also is a generalization of the one used by JPDA filter methods.

\section{MHT Methods and Track Repulsion Effect}
\label{sec:MHT-Tracking}

MHT methods aim at calculating the MAP sequence estimate of $\RV{x}_{0:k}$, $\RV{r}_{0:k}$, and $\RV{a}_{1:k}$ from $\V{z}_{1:k}$. This is done in two steps. First, the joint MAP sequence estimate of $\RV{r}_{0:k}$ and $\RV{a}_{1:k}$ is calculated, i.e.,
\be
\hspace{-3mm}\big(\hat{\V{r}}^{\text{MAP}}_{0:k}\!,\hat{\V{a}}^{\text{MAP}}_{1:k} \big) 
  =\ist \argmax_{\V{r}_{0:k},\ist\V{a}_{1:k}} \int \! f( \V{x}_{0:k}, \V{r}_{0:k}, \V{a}_{1:k} | \V{z}_{1:k} ) \, \text{d}\V{x}_{0:k} \ist. \!\! 
\label{eq:MHT0} 
\ee  
This is equivalent to searching for the PT-measurement association $(\V{r}_{0:k},\V{a}_{1:k})$ that is most probable given the measurements $\V{z}_{1:k}$ \cite{reid79}. The marginalization operation $\int \rmv\cdot\, \text{d}\V{x}_{0:k}$ can be performed sequentially and, assuming linear-Gaussian state-transition and measurement models, also in closed form. On the other hand, the number of  PT-measurement associations $(\V{r}_{0:k},\V{a}_{1:k} )$ that must be searched to perform the maximization \eqref{eq:MHT0} grows exponentially with $k$.

In the second step, the MAP estimate of $\RV{x}_{0:k}$ given $\V{r}_{0:k} \!=\! \hat{\V{r}}^{\text{MAP}}_{0:k}$ and $\V{a}_{1:k} \!=\! \hat{\V{a}}^{\text{MAP}}_{1:k}$ is calculated, i.e.,
\begin{align}
\hat{\V{x}}_{0:k}^{\text{MAP}} &= \ist\argmax_{\V{x}_{0:k}} f\big( \V{x}_{0:k} \big| \hat{\V{r}}^{\text{MAP}}_{0:k}\!,\hat{\V{a}}^{\text{MAP}}_{1:k} \!, \V{z}_{1:k}\big) \nn \\[.5mm]
&= \ist\argmax_{\V{x}_{0:k}} f\big( \V{x}_{0:k}, \hat{\V{r}}^{\text{MAP}}_{0:k}\!,\hat{\V{a}}^{\text{MAP}}_{1:k} \big| \V{z}_{1:k} \big)\ist.
\label{eq:MHT1}
\end{align}
Here, for $(\V{r}_{0:k} , \V{a}_{1:k}) \rmv=\rmv \big(\hat{\V{r}}^{\text{MAP}}_{0:k}\!,\hat{\V{a}}^{\text{MAP}}_{1:k} \big)$ fixed, 
\vspace{0.3mm}
the joint posterior $f\big( \V{x}_{0:k}, \hat{\V{r}}^{\text{MAP}}_{0:k}\!,\hat{\V{a}}^{\text{MAP}}_{1:k} \big| \V{z}_{1:k} \big)$ in \eqref{eq:jointPosteriorComplete1} factorizes into the posterior distributions $f\big( \V{x}_{0:k}^{(j)}, \hat{\V{r}}^{(j)\text{MAP}}_{0:k}\!,\hat{\V{a}}^{(j)\text{MAP}}_{1:k} \big| \V{z}_{1:k} \big)$, one for each PT $j$. The maximization \eqref{eq:MHT1} is then equivalent to a Bayesian smoothing operation---carried out by using, e.g., the Kalman smoother---for each PT $j$ in parallel with, and independently of, all the other PTs.

In a practical implementation, to reduce the computational complexity, the MAP estimates of the PT-measurement association sequence in \eqref{eq:MHT0} and of the kinematic PT state sequence in \eqref{eq:MHT1} are computed only over a sliding window of $N$ consecutive times steps. A hard decision on the PT-measurement association is made only at the oldest step within the window. The maximization in \eqref{eq:MHT0} and \eqref{eq:MHT1} is performed using the Auction algorithm or multiple frame assignment \cite{Ber:B91}.

MHT can be formulated in a hypothesis-oriented \cite{reid79} and a track-oriented \cite{Bar:B90,Bla04,Mor77} manner. The two formulations are equivalent under the assumption that target births and clutter follow Poisson point processes \cite{MorCho:C16}; however, the track-oriented MHT formulation offers a more compact representation of the \ac{da} problem. The computational complexity of the original hypothesis-oriented MHT formulation is still problematic due to the high number of hypotheses. However, it can be reduced by discarding unlikely hypotheses using, e.g., an efficient $m$-best assignment algorithm \cite{Cox96,DanNew06}. The more efficient track-oriented MHT methods \cite{Bar:B90,Bla04,Mor77} represent the \ac{da} hypotheses by a set of tree structures, where each tree represents the possible \ac{da} histories of a single PT. The most likely hypothesis is then found by choosing a leaf node from each single-PT tree so that no measurement is used by more than one PT. A fast hypothesis search is enabled by combinatorial optimization methods \cite{PatDeb92,deb97,PooRij93,PooGad06}. 

A limitation of MHT methods known as track repulsion can arise when targets come in close proximity. The estimated tracks then tend to have a larger distance than the true tracks. This effect is a consequence of performing hard \ac{da} by considering for PT state estimation (see \eqref{eq:MHT1}) only the single PT-measurement association $\big(\hat{\V{r}}^{\text{MAP}}_{0:k}\!,\hat{\V{a}}^{\text{MAP}}_{1:k} \big)$ in \eqref{eq:MHT0}. For an illustration, consider two targets that move on parallel tracks. We assume that the states of the targets include the targets' positions; furthermore, each target generates one measurement, which is the target's position plus Gaussian measurement noise. The detection probability is assumed to be one and there are no clutter measurements. The target tracks are supposed to be in close proximity in the sense that their distance is significantly smaller than the standard deviation of the measurement noise. The MAP decision rule \eqref{eq:MHT0} now always assigns to each measurement the nearest target. However, due to the small distance between the two targets, this association is incorrect if the measurement closest to a given target originated from the respective other target. A number of consecutive incorrect associations will then cause the position estimates in \eqref{eq:MHT1} to be more distant than the actual targets; this effect is known as track repulsion. The track repulsion effect is related to the use of MAP estimation and not to the choice of a specific optimization method or the number of hypotheses. It gets worse as the distance between the targets decreases.

\begin{figure}[t]
\centering

\begin{tikzpicture}
    \draw (0, 0) node[inner sep=0] {\includegraphics[scale=.241]{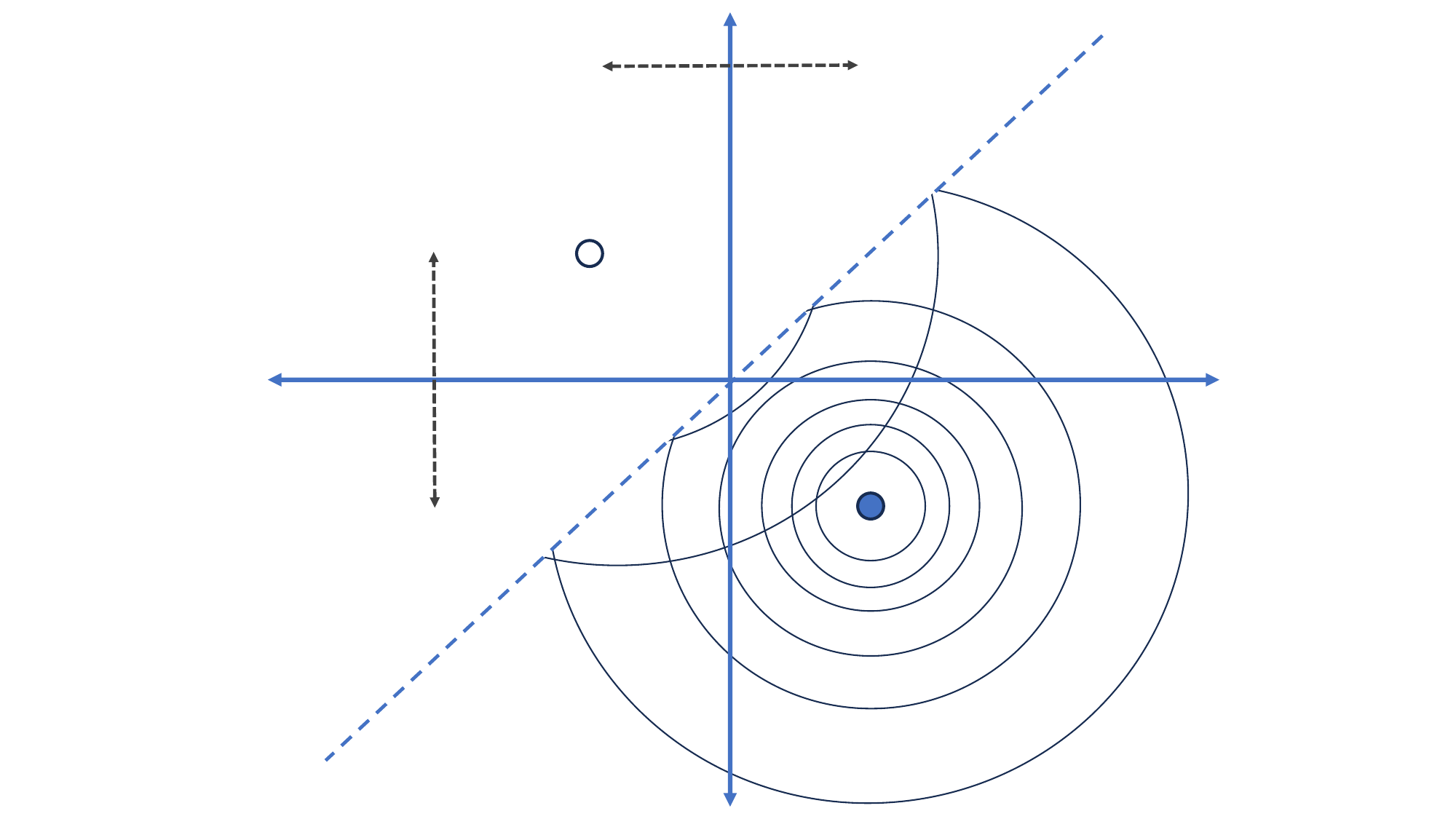}};
\draw (-.18, 1.47) node {\footnotesize $d$};    \draw (-1.83, .35) node {\footnotesize $d$};
    \draw (0.81,-.72) node {\footnotesize $\V{x}$};
        \draw (2.75,0) node {\footnotesize $x_1$};
    \draw (.3,2.15) node {\footnotesize $x_2$};
\end{tikzpicture}\vspace{1mm}
\caption{\label{fig:repulsion1} \rd{Equivalent 2-D geometry of the static 1-D example employing the ``hard association'' estimator $\hat{\RV{x}}$, which is $[\rv{z}^{(1)} \ist\ist\ist \rv{z}^{(2)}]^{\T}$ for $\rv{z}^{(1)} \geq \rv{z}^{(2)}$ and $[\rv{z}^{(2)} \ist\ist\ist \rv{z}^{(1)}]^{\T}$ for $\rv{z}^{(2)} < \rv{z}^{(1)}\rmv$. The level-curves visualize the distribution of $\hat{\RV{x}}$. The true state $\V{x}$ and its mirror image (mirrored on the $z_1=z_2$ line) are shown as a blue and a white bullet, respectively. When the 2-D joint measurement $[\rv{z}^{(1)} \ist\ist\ist \rv{z}^{(2)}]^{\T}$ crosses over the $z_1=z_2$ line, the level-curves are reflected, indicating the possibility of an incorrect hard association.}\vspace{-1mm}}
\vspace{-2mm}
\end{figure}

The reason why ``hard'' \ac{da} strategies such as those performed by MHT cause repulsion between close targets \rd{can be illustrated in a simple static one-dimensional (1-D) scenario. Consider two static 1-D target states $x^{(1)}=d/2$ and $x^{(2)}=-d/2$, separated by distance $d$. The targets generate the measurements  $\rv{z}^{(1)} \sim {\cal N}(x^{(1)},1) = {\cal N}(d/2,1)$ and  $\rv{z}^{(2)} \sim {\cal N}(x^{(2)},1) = {\cal N}(-d/2,1)$, which are assumed statistically independent. Fig.~\ref{fig:repulsion1} visualizes the underlying geometry in an equivalent 2-D state space representing the 2-D\linebreak 
joint target state $\V{x} \triangleq [x^{(1)} \ist\ist x^{(2)}]^{\T} \rmv= \big[ d/2 \hspace{1mm} -\rmv\rmv d/2\big]^{\T}\!$ and the distribution of the 2-D joint measurement $[\rv{z}^{(1)} \ist\ist \rv{z}^{(2)}]^{\T}\rmv$. We estimate the joint target state $\V{x}$ from the measurements $\rv{z}^{(1)}$ and $\rv{z}^{(2)}$ using a ``hard association'' estimator $\hat{\RV{x}} = [ \hat{\rv{x}}^{(1)} \ist\ist\ist \hat{\rv{x}}^{(2)} ]^{\T}$ defined as  $\hat{\RV{x}} = [\rv{z}^{(1)} \ist\ist\ist \rv{z}^{(2)}]^{\T}$ for $\rv{z}^{(1)} \geq \rv{z}^{(2)}$ and $\hat{\RV{x}} = [\rv{z}^{(2)} \ist\ist\ist \rv{z}^{(1)}]^{\T}$ for $\rv{z}^{(2)} < \rv{z}^{(1)}\rmv$. This estimator, which is also known as the global nearest neighbor (GNN) decision rule \cite{barShalom11}, employs a hard association strategy similar to MHT.}

\begin{figure}[t]
\centering

\begin{tikzpicture}
    \draw (0, 0) node[inner sep=0] {\includegraphics[scale=.241]{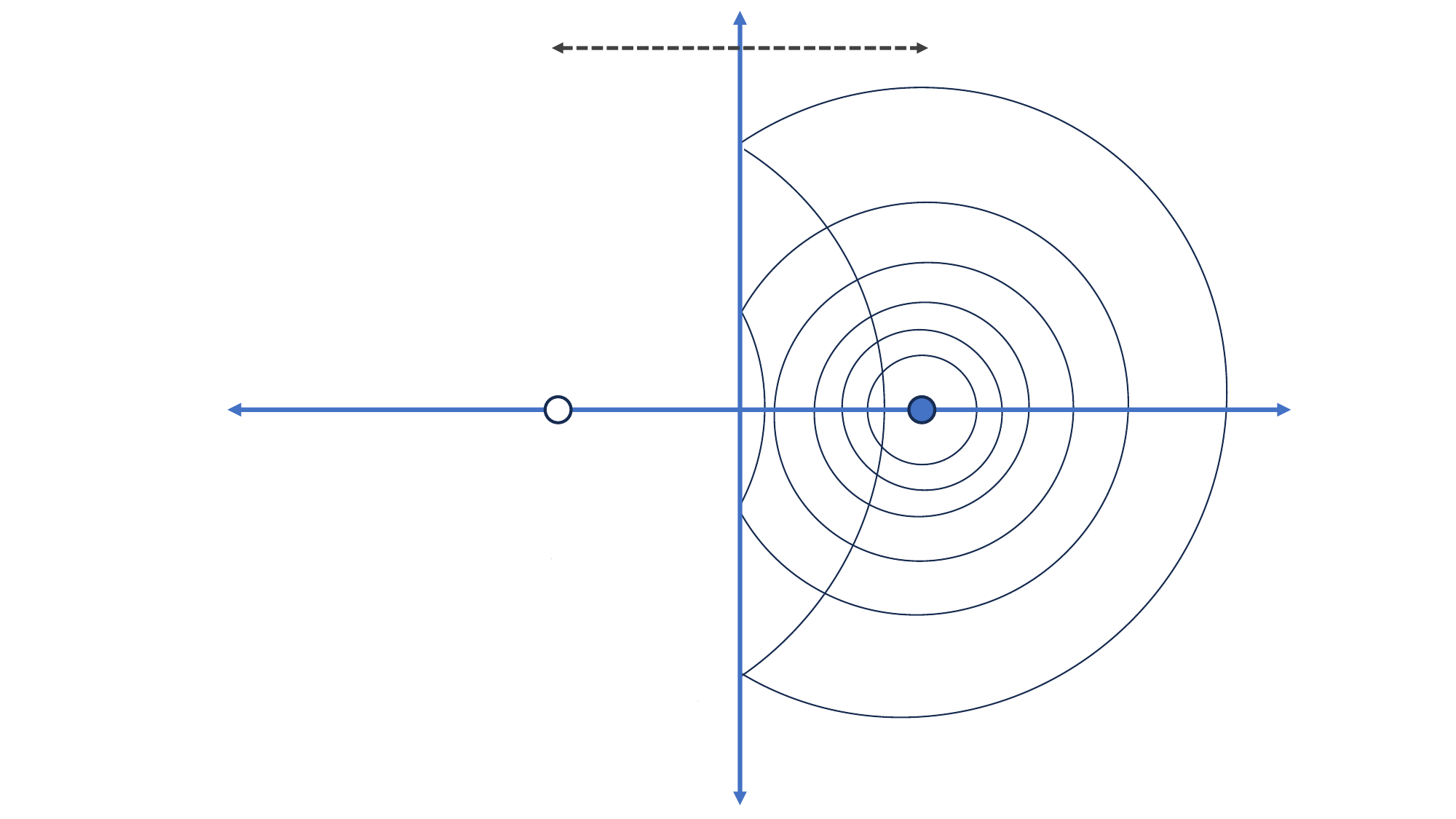}};
    \draw (-.35, 1.68) node {\footnotesize $\sqrt{2} d$};
    \draw (1.15,-.17) node {\footnotesize $\V{x}^{\mathrm{r}}$};
    \draw (3.15,-.2) node {\footnotesize $x^{\mathrm{r}}_1$};
    \draw (.35,2.22) node {\footnotesize $x^{\mathrm{r}}_2$};
\end{tikzpicture}
\vspace{1mm}
\caption{\label{fig:repulsion2} Same as Fig.~\ref{fig:repulsion1}, but $\V{x}$ rotated by $45^{\circ}\rmv$.}
\vspace{-2mm}
\end{figure}

\rd{To demonstrate track repulsion effects in the estimator $\hat{\RV{x}}$, we consider the expected difference between $\hat{\rv{x}}^{(1)}$ and $\hat{\rv{x}}^{(2)}\rmv$, i.e., $\mathrm{E}\{ \hat{\rv{x}}^{(1)} \!-\rmv \hat{\rv{x}}^{(2)} \}$. To this end, we first calculate $\mathrm{E}\{ \hat{\RV{x}} \}$. This calculation can be simplified by exploiting the rotational symmetry of the pdf of $[\rv{z}^{(1)} \ist\ist \rv{z}^{(2)}]^{\T}$ and rotating $\V{x} = \big[ d/2 \hspace{1mm} -\rmv\rmv d/2\big]^{\T}\rmv$ and $\hat{\RV{x}}$ by $45^{\circ}$ to obtain $\V{x}^{\mathrm{r}} \triangleq \big[d/\sqrt{2} \hspace{3mm} 0\big]^{\T}\!$ and $\hat{\RV{x}}^{\mathrm{r}}\rmv$, respectively. The result is shown in Fig.~\ref{fig:repulsion2}. The expected value of the rotated estimate $\hat{\RV{x}}^{\mathrm{r}}$ is given by $\mathrm{E}\{\hat{\RV{x}}^{\mathrm{r}}\} = \V{x}^{\mathrm{r}} + [\Delta^{\mathrm{r}} \hspace{1mm} 0 ]^{\T}$ with
\[
\Delta^{\mathrm{r}} = -2 \int_{-\infty}^0 \rmv\rmv y \ist  \frac{1}{\sqrt{2 \pi}} e^{-(y-\frac{d}{\sqrt{2}})^2/2} \ist \mathrm{d} y. 
\]
This can be rewritten as $\Delta^{\mathrm{r}} = \sqrt{\frac{2}{\pi}} \ist e^{-d^2/4} - \sqrt{2} \ist d \, Q\rmv\big(\frac{d}{\sqrt{2}}\big)$, where $Q(\alpha) \triangleq \frac{1}{\sqrt{2 \pi}} \int_{\alpha}^{\infty} \rmv\rmv  e^{-\zeta^2/2} \ist \mathrm{d} \zeta$ is the probability that a zero-mean Gaussian random variable with unit variance is larger than $\alpha$. After rotating back, we obtain the expectation of the original estimator as $\mathrm{E}\{\hat{\RV{x}}\} =  \big[ \mathrm{E}\{\hat{\rv{x}}^{(1)}\} \hspace{1.5mm} \ist\ist \mathrm{E}\{ \hat{\rv{x}}^{(2)} \} \big]^{\T}$ with $\mathrm{E}\{\hat{\rv{x}}^{(1)}\} = \frac{d}{2} + \Delta = x^{(1)} + \Delta$ and $\mathrm{E}\{\hat{\rv{x}}^{(2)}\} = -\frac{d}{2} - \Delta = x^{(2)} - \Delta$, where
\[
\Delta = \frac{\Delta^{\mathrm{r}}}{\sqrt{2}} = \frac{1}{\sqrt{\pi}} \ist e^{-d^2/4} - d \, Q \bigg( \frac{d}{\sqrt{2}} \bigg).
\]
Thus, the expected distance between $\hat{\rv{x}}^{(1)}$ and $\hat{\rv{x}}^{(2)}$ is 
\[
\mathrm{E}\{ \hat{\rv{x}}^{(1)} \!-\rmv \hat{\rv{x}}^{(2)} \} = \mathrm{E}\{\hat{\rv{x}}^{(1)}\} \rmv-\rmv \mathrm{E}\{\hat{\rv{x}}^{(2)}\} = d + 2 \ist \Delta \ist,
\]
which is seen to be larger by $2 \ist \Delta$ than the actual distance between the target states, $x^{(1)} \!-\rmv x^{(2)} \!=\rmv d$. We note that $\Delta$ is positive and increases with decreasing $d$. This analysis indicates that the considered hard association estimator leads to a repulsion of the target state estimates $\hat{\rv{x}}^{(1)}$ and $\hat{\rv{x}}^{(2)}$\rmv.}

The track repulsion effect can increase the estimation error and lead to track swaps (or identity switches) \cite{CorGriDet:C06,Willet07}. In particular, in a scenario with crossing tracks, the estimated tracks tend to ``bounce'' more often than they cross. A method for mitigating track repulsion was presented \vspace{1mm} in \cite{corcar12}.

\section{JPDA Filter Methods and Track Coalescence\\[-.5mm]Effect}
\label{sec:JPDA-Tracking}

The JPDA filter aims to calculate the MMSE estimates of the kinematic target states $\RV{x}_k^{(j)}\rmv$, i.e.,
\be
\hat{\V{x}}_k^{(j)\text{MMSE}} = \rmv \int\! \V{x}^{(j)}_k f\big(\V{x}^{(j)}_k \big|\V{z}_{1:k} \big)\, \text{d}\V{x}_k^{(j)} \rmv, \quad j \rmv=\rmv 1,\dots,L_k  \ist. 
\label{eq:JPDA_MMSE}
\ee
Here, $L_k$ denotes the number of actual targets (not PTs) at time $k$. The posterior pdf
\vspace{0.4mm}
of the joint kinematic target state $\RV{x}_k = \big[{\RV{x}}^{(1)\T}_k \rmv\cdots\ist {\RV{x}}^{(L_{k})\T}_k\big]^{\mathrm{T}}\!$ is given by
\begin{align}
f(\V{x}_k|\V{z}_{1:k}) &= \sum_{\V{a}_{k}}\ist  f(\V{x}_{k},\V{a}_{k}|\V{z}_{1:k}) \nn \\
&= \ist\sum_{\V{a}_{k}}  f(\V{x}_{k}|\V{a}_{k},\V{z}_{1:k}) \ist p(\V{a}_{k}|\V{z}_{1:k})\ist. \label{eq:JPDA_01} \\[-5mm]
\nn
\end{align}
This sum over all possible association events $\V{a}_{k}$ is also referred to as exhaustive averaging.

The JPDA filter is based on the approximation of the posterior \ac{da} pmf $p(\V{a}_{k}|\V{z}_{1:k})$ involved in \eqref{eq:JPDA_01} by the product of its marginals, i.e.\ \cite{barShalom11,mey18}
\be
p(\V{a}_{k}|\V{z}_{1:k}) \approx \prod_{j=1}^{L_k} p\big(a^{(j)}_{k} \big|\V{z}_{1:k}\big) \ist,
\label{eq:JPDA_marg}
\vspace{-2mm}
\ee
with
\vspace{.5mm}
\be
p\big(a^{(j)}_{k} \big|\V{z}_{1:k}\big) =\rmv \sum_{\V{a}_{k}^{(\sim j)}} \rmv p(\V{a}_{k}|\V{z}_{1:k}) \ist.
\label{eq:JPDA_marg2}
\vspace{-1mm}
\ee
Here, $\V{a}_{k}^{(\sim j)}$ denotes the vector $\V{a}_{k}$ with the $j$th entry $a^{(j)}_{k}\rmv$ removed. The posterior DA pmf $p(\V{a}_{k}|\V{z}_{1:k})$ involved in \eqref{eq:JPDA_01} and \eqref{eq:JPDA_marg2} is given as \cite{mey18}
\begin{equation}
\label{eq:JPDA_beta1}
p(\V{a}_{k}|\V{z}_{1:k}) \propto \prod_{j=1}^{L_k} w_k^{(j,a_k^{(j)})}\rmv, 
\vspace{-2.5mm}
\end{equation} 
with
\vspace{1mm}
\begin{equation}
\label{eq:JPDA_beta2}
w_k^{(j,a_k^{(j)})} \triangleq \rmv\int\ist  q\big(\V{x}^{(j)}_{k}\!, 1, a^{(j)}_{k}\rmv; \V{z}_{k} \big)\ist f\big(\V{x}_{k}^{(j)} \big|\V{z}_{1:k-1}\big)\ist \text{d}\V{x}_{k}^{(j)} \rmv, 
\vspace{1mm}
\end{equation} 
for $j = 1,\ldots,L_k$ and $a_k^{(j)} = 0,\ldots,M_k$. Here, the function $q\big(\V{x}^{(j)}_{k}\!, 1, a^{(j)}_{k}\rmv; \V{z}_{k} \big)$ is given by \eqref{eq:qFunction} with $\underline{\V{x}}_k^{(j)}$ replaced by $\V{x}_k^{(j)}$, and $ f\big(\V{x}_{k}^{(j)} \big|\V{z}_{1:k-1}\big)$ is the ``predicted'' posterior pdf of $\RV{x}_{k}^{(j)}\rmv$. Furthermore, based on the assumption that, for $a_k^{(j)} \rmv= m$, the measurement $\RV{z}_{k}^{(m)}$ is conditionally independent given $\V{x}_k^{(j)}$ of all past and future measurements $\RV{z}_{k'}^{(m')}$ and states $\RV{x}_{k'}^{(j)}$ with $k'\neq k$, it is shown in \cite{mey18} that the pdf $f(\V{x}_{k}|\V{a}_{k},\V{z}_{1:k})$ involved in \eqref{eq:JPDA_01} factorizes as 
\begin{equation}
\label{eq:JPDA_FACT}
f(\V{x}_{k}|\V{a}_{k},\V{z}_{1:k}) = \prod_{j=1}^{L_k} f\big(\V{x}_{k}^{(j)} \big|a_{k}^{(j)}\rmv,\V{z}_{1:k} \big)\ist.
\end{equation}
Then, by inserting \eqref{eq:JPDA_marg} into \eqref{eq:JPDA_01} and using the factorization \eqref{eq:JPDA_FACT}, we arrive at the approximate posterior pdf 
\begin{align}
f(\V{x}_k|\V{z}_{1:k}) \ist&\approx \sum_{\V{a}_{k}} \prod_{j=1}^{L_k} f\big(\V{x}_{k}^{(j)} \big|a_{k}^{(j)}\rmv,\V{z}_{1:k} \big) \ist p\big(a_{k}^{(j)} \big|\V{z}_{1:k}\big) \nn \\
 & = \prod_{j=1}^{L_k} \sum_{a_{k}^{(j)} = \ist 0}^{M_k} \!\rmv f\big(\V{x}_{k}^{(j)} \big|a_{k}^{(j)}\rmv,\V{z}_{1:k} \big) \ist p\big(a_{k}^{(j)} \big|\V{z}_{1:k}\big) \nn \\
 & = \prod_{j=1}^{L_k} \tilde{f}\big(\V{x}_{k}^{(j)} \big|\V{z}_{1:k}\big) \ist, \label{eq:JPDA_approx} \\[-8mm]
\nn
\end{align} 
with 
\be
\tilde{f}\big(\V{x}_{k}^{(j)} \big|\V{z}_{1:k}\big) \triangleq\! \sum_{a_{k}^{(j)} =\ist 0}^{M_k} \!\rmv f\big(\V{x}_{k}^{(j)} \big|a_{k}^{(j)}\rmv,\V{z}_{1:k} \big) \ist p\big(a_{k}^{(j)} \big|\V{z}_{1:k}\big)\ist. 
\label{eq:JPDA_approx2}
\ee
Here, the conditional \ac{pdf} $f\big(\V{x}_{k}^{(j)} \big|a_{k}^{(j)}\rmv,\V{z}_{1:k} \big)$ is given as follows~\cite{mey18}: for $a_{k}^{(j)}\!\rmv=\rmv m \rmv\in\rmv \{1,\ldots,M_k\}$, 
\begin{align}
&f\big(\V{x}_{k}^{(j)} \big|a_{k}^{(j)} \!\rmv=\rmv m,\V{z}_{1:k}\big) \nn\\[1mm]
&\hspace{-.5mm}=\rmv \frac{f\big(\V{z}_{k}^{(m)} \big|\V{x}_{k}^{(j)}\big) \ist f\big(\V{x}_{k}^{(j)} \big|\V{z}_{1:k-1}\big) }{
  \int f\big(\V{z}_{k}^{(m)} \big|\V{x}_{k}^{(j)'}\big) \ist f\big(\V{x}_{k}^{(j)'} \big|\V{z}_{1:k-1}\big) \ist \text{d}\V{x}_{k}^{(j)'}} \ist, \quad\!\! m \! \in\!  \{1,\ldots,M_k\} \ist, \nn\\[-3mm]
\label{eq:JPDA_approx3} \\[-7mm]
\nn
\end{align}
and for $a_{k}^{(j)}\!\rmv=\rmv 0$,
\begin{equation}
f\big(\V{x}_{k}^{(j)} \big|a_{k}^{(j)} \!\rmv=\rmv 0,\V{z}_{1:k} \big) \rmv=\rmv f\big(\V{x}_{k}^{(j)} \big|\V{z}_{1:k-1} \big) \ist.  
\label{eq:JPDA_approx3a}
\end{equation}
We note that, based on the approximation \eqref{eq:JPDA_marg}, the pdfs $\tilde{f}\big(\V{x}_{k}^{(j)} \big|\V{z}_{1:k}\big)$ in \eqref{eq:JPDA_approx} and \eqref{eq:JPDA_approx2} approximate the marginal posterior pdfs $f\big(\V{x}_{k}^{(j)} \big|\V{z}_{1:k}\big)$. An important, often overlooked property of the JPDA solution in \eqref{eq:JPDA_approx2} is that if we have two identical predicted pdfs $f\big(\V{x}_{k}^{(j)} \big|\V{z}_{1:k-1} \big)$---which enter \eqref{eq:JPDA_approx2} via \eqref{eq:JPDA_approx3} and \eqref{eq:JPDA_approx3a}---for two different targets $j \neq j'$, then we also have identical approximate marginal posterior pdfs \cite{barShalom11,mey18}, i.e.,
\begin{align}
&f\big(\V{x}_{k}^{(j')} \big|\V{z}_{1:k-1}\big) = f\big(\V{x}_{k}^{(j)} \big|\V{z}_{1:k-1}\big) \nn\\[1mm]
&\hspace{18mm}\Rightarrow \tilde{f}\big(\V{x}_{k}^{(j')} \big|\V{z}_{1:k}\big) = \tilde{f}\big(\V{x}_{k}^{(j)} \big|\V{z}_{1:k}\big). \label{eq:ast}
\end{align}
This is due to symmetry in the computation of \eqref{eq:JPDA_marg2} and \eqref{eq:JPDA_approx2} and can easily be verified.

In conventional JPDA filter methods, differently from MHT methods, the sequence of existence indicators $\V{r}_{0:k}$, and thus also the number of targets $L_k$, are assumed known. In practical implementations, a heuristic for track initialization and termination provides an estimate of $\V{r}_{0:k}$. For linear and Gaussian measurement and motion models and a Gaussian prior for the kinematic states $\RV{x}_{k}^{(j)}\rmv$, as assumed in the original formulation of the JPDA filter \cite{barShalom74,fortmann83,barShalom11}, the approximate marginal posterior pdfs $\tilde{f}\big(\V{x}^{(j)}_k \big|\V{z}_{1:k}\big)$ in \eqref{eq:JPDA_approx2} are Gaussian mixture pdfs, which are further approximated by Gaussian pdfs.

A deficiency of the JPDA filter is the track coalescence effect: when targets come close to each other, the estimated tracks tend to merge and become indistinguishable. This behavior is due to the facts that (i) the predicted marginal posterior pdfs $f\big(\V{x}_{k}^{(j)} \big|\V{z}_{1:k-1}\big)$ involved in \eqref{eq:JPDA_approx3} and \eqref{eq:JPDA_approx3a} become similar when the targets come close to each other, and (ii) the approximate marginal posterior pdfs $\tilde{f}\big(\V{x}^{(j)}_k \big|\V{z}_{1:k}\big)$ are calculated via expression \eqref{eq:JPDA_approx2}. 
More specifically, performing ``soft \ac{da}'' by means of the summation over all possible associations $a_{k}^{(j)} \!\rmv\in\rmv \{0,1,\ldots,M_k\}$, as done in \eqref{eq:JPDA_approx2}, has the effect that targets $j$ with similar predicted marginal posterior pdfs $f\big(\V{x}_{k}^{(j)} \big|\V{z}_{1:k-1}\big)$---which enter \eqref{eq:JPDA_approx2} via \eqref{eq:JPDA_approx3} and \eqref{eq:JPDA_approx3a}---tend to also have similar approximate marginal posterior pdfs $\tilde{f}\big(\V{x}^{(j)}_k \big|\V{z}_{1:k}\big)$ (cf.~\eqref{eq:ast}), which is due to symmetry in the computations \eqref{eq:JPDA_marg2} and \eqref{eq:JPDA_approx2}. This, in turn, results in similar MMSE state estimates according to \eqref{eq:JPDA_MMSE} (with $f\big(\V{x}^{(j)}_k \big|\V{z}_{1:k}\big)$ replaced by $\tilde{f}\big(\V{x}^{(j)}_k \big|\V{z}_{1:k}\big)$). A detailed analysis of the track coalescence effect will be presented in Section \ref{sec:closerlook}.

Variants and extensions of the JPDA filter include the JIPDA filter \cite{musicki04,ChaMor11}, the JPDA* filter \cite{bloblo00}, and the set JPDA (SJPDA) filter \cite{SveSve11}. The JIPDA filter extends the JPDA filter by a binary existence indicator to systematically account for target existence, but suffers from track coalescence just as the conventional JPDA filter. The JPDA* filter mitigates track coalescence by pruning target-measurement associations. However, this comes at the cost of a reduced tracking performance in more challenging tracking scenarios with a high number of clutter measurements and missed detections. The SJPDA filter is based on the \ac{ospa} estimator \cite{SveSve11}, which is in turn based on the minimization of the mean OSPA metric \cite{schuhmacher08}. Similar to the JPDA* filter, the SJPDA filter exhibits reduced track coalescence effects. Efficient implementations are based on convex optimization techniques \cite{Cro13}. To maintain track continuity, the SJPDA filter relies on post-processing techniques \cite{Cro11}, which are not required for the other JPDA filter variants or MHT methods.

A potential limitation of the JPDA filter and its variants is the fact that their complexity scales exponentially with the number of targets and the number of measurements, which is due to the marginalization operation in \eqref{eq:JPDA_marg2} or equivalent marginalization operations. Gating and clustering strategies can significantly reduce the complexity in many cases, but fail when many objects come in close proximity. 
A potential solution in such scenarios is to switch to JPDA-type tracking methods of lower complexity such as the methods proposed in \cite{MusLaS08} and \cite{Fit90}. Other solutions rely on the pruning of target-measurement associations---which is inherently done in the JPDA* filter---or exploit the potential independence of target-measurement associations \cite{horridge06}.

\vspace{-1mm}
\section{BP Method}
\label{sec:BP-Tracking}

BP-based MTT methods \cite{mey17,mey18,SolMeyBraHla:J19,MeyWil:J21,Gag20,KroMeyHla:J20,MeyGem:J21,Gag21Fus} aim at computing the marginal posterior pdf/pmf $f\big(\V{x}^{(j)}_{k}\rmv, r^{(j)}_{k}\big|\V{z}_{1:k}\big)$ for each PT $j \in \{1,\dots,J_k\}$. This marginal posterior pdf/pmf is then used to perform target detection and MMSE state estimation. In what follows, we will consider the specific BP method introduced in \cite{mey18}. For target detection, the marginal posterior pmf $p\big(r^{(j)}_{k} \big| \V{z}_{1:k} \big)$ of the existence indicator $\rv{r}^{(j)}_{k}\rmv$ is obtained from $f\big(\V{x}^{(j)}_{k}\rmv, r^{(j)}_{k}\big|\V{z}_{1:k}\big)$ as
\begin{equation*}
p\big(r^{(j)}_{k} \big| \V{z}_{1:k} \big) = \int \rmv f\big(\V{x}^{(j)}_{k}\rmv,r^{(j)}_{k} \big| \V{z}_{1:k}  \big) \ist \mathrm{d} \V{x}^{(j)}_{k}. 
\vspace{-1mm}
\end{equation*}
PT $j$ is then detected---i.e., declared to exist---if $p\big(r^{(j)}_{k} \!\rmv=\!1\big|\V{z}_{1:k})$ is larger than a predefined threshold $P_{\text{th}}$. Next, for all PTs that are declared to exist, MMSE state estimation is performed according to
\begin{equation}
\label{eq:BP-MMSE}
\hat{\V{x}}^{(j)\text{MMSE}}_{k} \ist = \rmv\int \rmv \V{x}^{(j)}_k f\big(\V{x}^{(j)}_{k} \big| r^{(j)}_{k} \!\rmv=\! 1, \V{z}_{1:k} \big) \ist \mathrm{d}\V{x}^{(j)}_k \rmv,
\vspace{-.5mm}
\end{equation}
with
\begin{equation}
\label{eq:BP-post}
f\big(\V{x}^{(j)}_{k} \big| r^{(j)}_{k} \!\rmv=\! 1, \V{z}_{1:k} \big) = \frac{f\big(\V{x}^{(j)}_{k}\rmv,r^{(j)}_{k} \!\rmv=\! 1 \big| \V{z}_{1:k} \big)}{p\big(r^{(j)}_{k} \!\rmv=\! 1 \big| \V{z}_{1:k} \big)}\ist.
\end{equation}

It remains to calculate $f\big(\V{x}^{(j)}_{k}\rmv, r^{(j)}_{k}\big|\V{z}_{1:k}\big)$. We have
\[
f\big(\V{x}^{(j)}_k\rmv, r^{(j)}_{k}\big| \V{z}_{1:k}\big) =\rmv \int \! \sum_{ \V{r}^{(\sim j)}_{k}} \rmv f(\V{x}_k, \V{r}_{k} | \V{z}_{1:k}) \ist \mathrm{d} \V{x}^{(\sim j)}_{k} \rmv, 
\vspace{-1.5mm}
\]
where $\V{r}^{(\sim j)}_{k}\rmv$ denotes $\V{r}_{k}$ with the entry ${r}^{(j)}_{k}\rmv$ removed, $\V{x}^{(\sim j)}_{k}\rmv$ denotes $\V{x}_{k}$ with the subvector $\V{x}^{(j)}_{k}\rmv$ removed, and $f(\V{x}_{k}, \V{r}_{k} | \V{z}_{1:k})$ is a marginal pdf/pmf of the joint posterior \ac{pdf}/\ac{pmf} $f( \V{x}_{0:k}, \V{r}_{0:k}, \V{a}_{1:k}, \V{b}_{1:k} | \V{z}_{1:k} )$, i.e.,
\begin{align}
&\hspace{-3mm} f(\V{x}_{k},\V{r}_{k} | \V{z}_{1:k}) \nn \\[1.5mm]
&\hspace{-4mm} =\rmv \int \!\!\rmv \sum_{\V{r}_{0:k-1}} \sum_{\V{a}_{1:k}} \ist \sum_{\V{b}_{1:k}} \ist f( \V{x}_{0:k}, \V{r}_{0:k}, \V{a}_{1:k}, \V{b}_{1:k} | \V{z}_{1:k} ) \, \mathrm{d} \V{x}_{0:k-1} \ist . 
\!\! \label{eq:marginalization1}
\end{align}
By inserting the factorization \eqref{eq:jointPosteriorComplete}, \eqref{eq:marginalization2_1} of $f( \V{x}_{0:k}, \V{r}_{0:k}, \V{a}_{1:k},\linebreak 
\V{b}_{1:k} | \V{z}_{1:k} )$ and carrying out the marginalizations with respect to $\V{x}_{0:k-2}$, $\V{r}_{0:k-2}$, $\V{a}_{1:k-1}$, and $\V{b}_{1:k-1}$, it can be shown that \cite{mey18}
\begin{align}
f(\V{x}_k, \V{r}_{k} | \V{z}_{1:k}) &=\rmv \int \rmv \sum_{\V{r}_{k-1}} \sum_{\V{a}_{k}} \sum_{\V{b}_{k}} 
  f( \V{x}_{k-1}, \V{r}_{k-1} | \V{z}_{1:k-1} ) \nn\\[0mm]
&\hspace{4mm} \times g(\V{x}_{k}, \V{r}_{k}, \V{a}_{k}, \V{b}_{k}, \V{x}_{k-1}, \V{r}_{k-1};\V{z}_{k}) \ist\ist \mathrm{d} \V{x}_{k-1}, \nn \\
 \label{eq:marginalization2}  \\[-7.5mm]
\nn
\end{align}
where $g(\V{x}_{k}, \V{r}_{k}, \V{a}_{k}, \V{b}_{k}, \V{x}_{k-1}, \V{r}_{k-1};\V{z}_{k})$ was defined in \eqref{eq:marginalization2_1}.

Two observations can be made at this point. First, it can be concluded from \eqref{eq:BP-MMSE}--\eqref{eq:marginalization2} that the MMSE estimator $\hat{\V{x}}^{(j)\text{MMSE}}_{k}$ in \eqref{eq:BP-MMSE} is a ``single-scan solution'' in the sense that, as shown by \eqref{eq:marginalization2}, the marginal posterior \ac{pdf}/\ac{pmf} $f(\V{x}_k, \V{r}_{k} | \V{z}_{1:k})$ at time $k$ can be directly obtained from the marginal posterior \ac{pdf}/\ac{pmf} at time $k \!-\! 1$,\linebreak $f(\V{x}_{k-1}, \V{r}_{k-1} | \V{z}_{1:k-1})$, and the current measurement $\V{z}_{k}$ (which enters via $g(\V{x}_{k}, \V{r}_{k}, \V{a}_{k}, \V{b}_{k}, \V{x}_{k-1}, \V{r}_{k-1};\V{z}_{k})$). In other words, $f(\V{x}_{k-1}, \V{r}_{k-1} | \V{z}_{1:k-1})$ subsumes and provides all the relevant information from past time steps. We note that a single-scan method based on MMSE estimation does not necessarily perform worse than a multiscan method based on MAP estimation (such as an MHT method).

Second, the computational complexity of evaluating the expressions \eqref{eq:marginalization2} and \eqref{eq:marginalization2_1} is much smaller than that of directly performing the marginalization in \eqref{eq:marginalization1}. Indeed, the summation over $\V{r}_{0:k-1}$ in \eqref{eq:marginalization1} involves $2^{J_{0}}\cdots 2^{J_{k-1}} = 2^{ \hspace{.1mm} \sum^{k-1}_{k'=0} J_{k'} }\rmv$ terms, whereas the repeated application of \eqref{eq:marginalization2} from $0$ to $k-1$ only involves $\sum^{k-1}_{k'=0} \ist 2^{\hspace{.1mm} J_{k'}}$ terms. This complexity reduction is a consequence of the temporal factorization structure of the MTT problem. However, we still have to perform the marginalizations in \eqref{eq:marginalization2}, whose complexity scales exponentially in the number of PTs and the number of measurements. These marginalizations can be computed in an efficient (though approximate) manner using the BP approach described in the following, which has only a linear complexity scaling.

The BP method operates on a factor graph representing the statistical model of the considered estimation problem \cite{kschischang01,koller09}. The factor graph for our statistical model is shown in Fig.~\ref{fig:factorGraph}. The marginal pdfs/pmfs $f\big(\V{x}^{(j)}_{k} \rmv,\V{r}^{(j)}_{k}\big|\V{z}_{1:k}\big)$, $j = 1,\dots,J_k$ needed for target detection and state estimation as discussed above are calculated efficiently by performing local operations corresponding to the individual graph nodes and exchanging the results of these local operations---called ``messages''---along the graph edges \cite{mey17,mey18}. \rd{In the following, we outline the calculation of these messages and the resulting beliefs (see also Fig.~\ref{fig:factorGraph}). More details are provided in \cite{mey18}, and a particle-based implementation is described \vspace{1.5mm} in \cite{mey17}.  
\begin{enumerate}
\item \emph{Prediction}: 
First, state prediction is performed separately for each legacy PT $j \rmv\in\rmv \{1,\ldots,J_{k-1}\}$ by sending a message $\alpha_k^{(j)}\big(\underline{\V{x}}_{k}^{(j)}\rmv, \underline{r}_k^{(j)}\big)$ from factor node ``$f\big( \underline{\V{x}}_{k}^{(j)}\rmv, \underline{r}_{k}^{(j)} \big| \V{x}_{k-1}^{(j)}, r_{k-1}^{(j)} \big)$'' to variable node ``$\big[\underline{\V{x}}^{(j)\T}_k\ist\ist \underline{r}^{(j)}_k\big]^{\T}$''. This message is calculated according to 
\begin{align*}
\alpha_k^{(j)}\big(\underline{\V{x}}_{k}^{(j)}\rmv, \underline{r}_k^{(j)}\big) &= \!\rmv \sum_{r_{k-1}^{(j)} \in \{0,1\}}\int \! 
  f\big( \underline{\V{x}}_{k}^{(j)}\rmv, \underline{r}_{k}^{(j)} \big| \V{x}_{k-1}^{(j)}, r_{k-1}^{(j)}\big) \\
&\hspace{4mm} \times  f\big(\V{x}_{k-1}^{(j)}, r_{k-1}^{(j)} \big| \V{z}_{1:k-1}\big) \, \mathrm{d}\V{x}_{k-1}^{(j)} \ist.\nn \\[-5mm] \nn
\end{align*}
Here, $f\big(\V{x}_{k-1}^{(j)}, r_{k-1}^{(j)} \big| \V{z}_{1:k-1}\big)$ is the marginal posterior pdf/pmf of legacy PT $j \in \{1,\ldots,J_{k-1}\}$ at 
\vspace{1.5mm}
time $k-1$.
\item \emph{Measurement Evaluation}: 
Next, for each legacy PT $j \in \{1,\ldots,J_{k-1}\}$, the following message is passed from factor node ``$q\big( \underline{\V{x}}^{(j)}_{k}\rmv, \underline{r}^{(j)}_{k}\rmv, a_{k}^{(j)};\V{z}_{k} \big)$'' to variable node ``$a_k^{(j)}$'':
\begin{align*}
&\beta_k^{(j)}\big(a_k^{(j)}\big) \\[1mm]
&\rmv= \!\rmv \sum_{\underline{r}^{(j)}_{k} \in \{0,1\}} \int \rmv\rmv q\big( \underline{\V{x}}^{(j)}_{k}\rmv, \underline{r}^{(j)}_{k}\rmv, a_{k}^{(j)};\V{z}_{k} \big) 
\ist \alpha_k^{(j)}\big(\underline{\V{x}}_{k}^{(j)}\rmv, \underline{r}_k^{(j)}\big) \, \mathrm{d}\underline{\V{x}}^{(j)}_{k}\!. \\[-5.5mm]
\end{align*}
Similarly, for each new PT $m \in \{1,\ldots,M_k\}$, the following message is passed from factor node ``$v\big( \overline{\V{x}}^{(m)}_{k} \rmv,\linebreak 
\overline{r}^{(m)}_{k}\rmv,b_{k}^{(m)};\V{z}^{(m)}_{k} \big)$'' to variable node ``$b_k^{(m)}$'':
\begin{align*}
&\xi_k^{(m)}\big(b^{(m)}_{k}\big)  \\[0mm]
&\hspace{2mm}= \!\rmv \sum_{\overline{r}^{(m)}_{k} \in \{0,1\}} \int \rmv\rmv v\big( \overline{\V{x}}^{(m)}_{k}\rmv, \overline{r}^{(m)}_{k}\rmv, b^{(m)}_{k} ; \V{z}^{(m)}_{k} \big) \ist \mathrm{d}\overline{\V{x}}^{(m)}_{k}\!.
\end{align*}
\item \emph{Probabilistic DA}: 
Probabilistic DA is now performed iteratively by sending, at message passing iteration $p \in \{ 1,\ldots,P \}$, a message $\varphi_{k}^{[p](j\rightarrow m)}$ from variable node ``$a_k^{(j)}$'' via factor node ``$\Psi_{j,m}\big(a^{(j)}_{k} \rmv,b^{(m)}_{k}\big)$'' to variable node ``$b_k^{(m)}$'' and a message $\nu_{k}^{[p](m\rightarrow j)}$ from variable node ``$b_k^{(m)}$'' via factor node ``$\Psi_{j,m}\big(a^{(j)}_{k} \rmv,b^{(m)}_{k}\big)$'' to variable node ``$a_k^{(j)}$''. These messages are calculated according
\vspace{-2mm} 
to
\begin{align*}
\varphi_{k}^{[p](j\rightarrow m)} &= \frac{\beta_{k}^{(j)}(m)}
{\beta_{k}^{(j)}(0) + \sum_{\substack{m'=1\\ m'\neq m}}^{M_{k}} \rmv \beta_{k}^{(j)}(m') \ist \nu_{k}^{[p](m'\rightarrow j)}} \\[1mm]
\nu_{k}^{[p](m\rightarrow j)} &=
\frac{\xi_{k}^{(m)}(j)}
{\xi_{k}^{(m)}(0) + \sum_{\substack{ j'=1\\ j'\neq j}}^{J_{k-1}} \xi_{k}^{(m)}(j') \ist
  \varphi_{k}^{[p-1](j'\rightarrow m)}} \ist, \\[-6mm] 
\end{align*}
for $j \rmv=\rmv 1,\ldots,J_{k-1}$ and $m \rmv=\rmv 1,\ldots,M_k$. 
This iterative procedure is initialized with $\varphi_{k}^{[0](j\rightarrow m)} \!=\! 1$. After the last iteration $p \rmv=\rmv P$, a message $\kappa^{(j)}_{k}\big(a^{(j)}_{k}\big)$ is passed from variable node ``$a^{(j)}_{k}$'' to factor node ``$q\big( \underline{\V{x}}^{(j)}_{k} \rmv,\underline{r}^{(j)}_{k} \rmv,a_{k}^{(j)};\V{z}_{k} \big)$''. This message is given for $a^{(j)}_{k} = m \in \{0,1,\ldots,M_k\}$ 
\vspace{0mm}
by
\[
\kappa^{(j)}_{k}(m) = \!\frac{\beta_{k}^{(j)}(m) \ist \nu_{k}^{[P](m\rightarrow j)}}
{\beta_{k}^{(j)}(0) + \sum_{m'=1}^{M_{k}}\beta_{k}^{(j)}(m')\ist \nu_{k}^{[P](m'\rightarrow j)}} \ist. 
\vspace{.5mm}
\]
Similarly, a message $\iota^{(m)}_{k}\big(b^{(m)}_{k} \big)$ is passed from variable node ``$b^{(m)}_{k}$'' to factor node ``$v\big( \overline{\V{x}}^{(m)}_{k}\rmv,\overline{r}^{(m)}_{k} \rmv, b_{k}^{(m)};\V{z}_{k}^{(m)} \big)$''. It is given for $b^{(m)}_{k} = j \in \{0,1,\ldots,J_{k-1}\}$ 
by 
\[
\iota^{(m)}_{k}(j) = \! \frac{\xi_k^{(m)}(j) \ist \varphi_{k}^{[P](j\rightarrow m)}}
{ \xi_k^{(m)}(0) + \sum_{j'=1}^{J_{k-1}} \xi_k^{(m)} (j') \ist\varphi_{k}^{[P](j'\rightarrow m)}}\ist.
\vspace{1.5mm}
\]
\item \emph{Measurement Update}: 
Measurement update steps are next applied for the legacy PTs and the new PTs. For each legacy PT $j \rmv\in\rmv \{1,\ldots,J_{k-1}\}$, the message
\[
\gamma^{(j)}_k\big(\underline{\V{x}}_{k}^{(j)}\rmv,\underline{r}_k^{(j)}\big) =\! \sum_{a_k^{(j)}=\ist 0}^{M_k} \!\rmv q\big( \underline{\V{x}}^{(j)}_{k}\rmv, \underline{r}^{(j)}_{k}\rmv, a_k^{(j)} ; \V{z}_{k} \big) \ist \kappa^{(j)}_{k}\big(a_k^{(j)}\big)
\vspace{-1mm}
\]
is passed from factor node ``$q\big( \underline{\V{x}}^{(j)}_{k}\rmv, \underline{r}^{(j)}_{k}\rmv, a^{(j)}_{k} ; \V{z}_{k} \big)$'' to variable node 
``$\big[\underline{\V{x}}^{(j)\T}_k\ist\ist \underline{r}^{(j)}_k\big]^{\T}$''. Furthermore, for each new PT $m\rmv\in\rmv\{1,\ldots,M_k\}$, the message 
\[
\varsigma_k^{(m)}\big(\overline{\V{x}}^{(m)}_{k}\rmv,\overline{r}_k^{(m)} \big) 
  = v\big( \overline{\V{x}}^{(m)}_{k}\rmv, \overline{r}^{(m)}_{k}\rmv, b^{(m)}_{k} ; \V{z}^{(m)}_{k} \big) \ist \iota^{(m)}_{k}\big(0 \big)
\]
is passed from factor node ``$v\big( \overline{\V{x}}^{(m)}_{k}\rmv, \overline{r}^{(m)}_{k}\rmv, b^{(m)}_{k} ; \V{z}^{(m)}_{k} \big)$'' to variable node \vspace{1.5mm} ``$\big[\overline{\V{x}}^{(m)\T}_k\ist\ist\ist \overline{r}^{(m)}_k\big]^{\T}$''.
\item \emph{Belief calculation}:
Finally, for each legacy PT $j\in\{1,\ldots,J_{k-1}\}$, a belief $\tilde{f}\big(\underline{\V{x}}^{(j)}_{k}\rmv,\underline{r}^{(j)}_{k}\big)$ approximating the marginal posterior pdf/pmf $f\big(\underline{\V{x}}^{(j)}_{k}\rmv,\underline{r}^{(j)}_{k} \big| \V{z}_{1:k} \big)$ is obtained as
\vspace{0mm}
\[
\tilde{f}\big(\underline{\V{x}}^{(j)}_{k}\rmv,\underline{r}_k^{(j)}\big) 
  = \frac{1}{\underline{C}^{(j)}_{k}} \ist\ist \alpha^{(j)}_k\big(\underline{\V{x}}^{(j)}_{k}\rmv,\underline{r}_k^{(j)}\big) \ist 
  \gamma^{(j)}_k\big(\underline{\V{x}}^{(j)}_{k}\rmv,\underline{r}_k^{(j)}\big) , 
\vspace{-1mm}
\]
with $\underline{C}^{(j)}_{k} \!\triangleq \sum_{\underline{r}_{k}^{(j)} \in \{0,1\}}\int \alpha^{(j)}_k\big(\underline{\V{x}}^{(j)}_{k}\rmv,\underline{r}_k^{(j)}\big) \ist \gamma^{(j)}_k\big(\underline{\V{x}}^{(j)}_{k}\rmv,\underline{r}_k^{(j)}\big)\linebreak 
\times \mathrm{d} \underline{\V{x}}^{(j)}_{k}\!$. Analogously, for each new PT $m\in\{1,\ldots,\linebreak 
M_k\}$, a belief $\tilde{f}\big(\overline{\V{x}}^{(m)}_{k}\rmv,\overline{r}^{(m)}_{k}\big)$ approximating the marginal posterior pdf/pmf $f\big(\overline{\V{x}}^{(m)}_{k}\rmv,$ $\overline{r}^{(m)}_{k} \big| \V{z}_{1:k} \big)$ is obtained as
\[
\tilde{f}\big(\overline{\V{x}}^{\ist(m)}_{k}\rmv,\overline{r}_k^{(m)}\big) 
  = \frac{1}{\overline{C}^{(m)}_{k}} \ist\ist \varsigma_k^{(m)}\big(\overline{\V{x}}^{(m)}_{k}\rmv,\overline{r}_k^{(m)}\big) ,
\vspace{-1mm}
\]
with $\overline{C}^{(m)}_{k} \!\triangleq\rmv \sum_{\overline{r}_{k}^{(j)} \in \{0,1\}} \int \varsigma_k^{(m)}\big(\overline{\V{x}}^{(m)}_{k}\rmv,\overline{r}_{k}^{(j)}\big) \ist \mathrm{d} \overline{\V{x}}^{(m)}_{k}\!$. 
\vspace{2mm}
\end{enumerate}}

This BP method systematically exploits the conditional independence structure of the involved random variables, as expressed by the factorization structure of \eqref{eq:jointPosteriorComplete}, \eqref{eq:marginalization2_1} and  represented by the factor graph of Fig.~\ref{fig:factorGraph}, for a large reduction of computational complexity. Here, in particular, the spatial factorization structure across PTs expressed by \eqref{eq:jointPosteriorComplete} and \eqref{eq:marginalization2_1} is exploited in addition to the temporal factorization structure. Since the factor graph in Fig.~\ref{fig:factorGraph} has loops, \rd{the beliefs $\tilde{f}\big(\underline{\V{x}}^{(j)}_{k}\rmv,\underline{r}_k^{(j)}\big)$ and $\tilde{f}\big(\overline{\V{x}}^{\ist(m)}_{k}\rmv,\overline{r}_k^{(m)}\big)$ only provide} approximations to the marginal posterior pdfs/pmfs $f\big(\underline{\V{x}}^{(j)}_{k}\rmv,\underline{r}_k^{(j)} \big|\V{z}_{1:k}\big)$ and $f\big( \overline{\V{x}}^{(m)}_{k}\rmv,\overline{r}_k^{(m)} \big|\V{z}_{1:k}\big)$, respectively \cite{kschischang01}. In the case of our factor graph, these approximations are sufficiently accurate to yield excellent tracking performance. However, the loops in the factor graph also lead to overconfident beliefs \cite{WeiFre:01}, i.e., the spread of the beliefs underestimates the true posterior uncertainty of the respective random variables.

Since the BP method performs a marginalization similar to JPDA filter methods, one may suspect that it suffers from track coalescence to a similar extent. Surprisingly, this is not the case: as analyzed in the next section and further evidenced by our simulation results in Section \ref{sec:results}, the BP method exhibits track coalescence to a lesser extent than traditional JPDA filtering methods. Moreover, it does not suffer from track repulsion effects.

\vspace{-1mm}
\section{A Closer Look at the Track Coalescence Effect}
\label{sec:closerlook}

In this section, we take a closer look at the track coalescence effect and investigate why BP methods exhibit this effect to a lesser extent than JPDA filter methods.

\vspace{-1mm} 

\subsection{JPDA Filter}
\label{sec:closerlook_jpda}

Let us first reconsider the JPDA filter from Section \ref{sec:JPDA-Tracking} for the case where the number of targets is fixed and known, i.e., $L_k \rmv=\rmv L$ for all times $k$. \rd{Here, using \eqref{eq:JPDA_beta1} and \eqref{eq:JPDA_beta2},} the posterior \ac{da} \ac{pmf} $p(\V{a}_{k}|\V{z}_{1:k})$ in \eqref{eq:JPDA_marg} can be expressed as
\begin{align}
p(\V{a}_{k}|\V{z}_{1:k}) &\propto  \prod^{L}_{j = 1} \int q\big(\V{x}^{(j)}_{k}\!, 1, a^{(j)}_{k}\rmv; \V{z}_{k} \big)\ist f\big(\V{x}^{(j)}_{k} \big| \V{z}_{1:k-1}\big) \ist \mathrm{d} \V{x}_{k}^{(j)}. \nn \\[-2.5mm]
\label{eq:jointAssoc} \\[-6mm]
\nn
\end{align}
If $f\big(\V{x}^{(j)}_{k} \big| \V{z}_{1:k-1}\big)$, i.e., the ``prior information'' about $\RV{x}^{(j)}_{k}$ at time $k$, is equal for all $j \rmv=\rmv 1,\dots,L$, then $f\big(\V{x}_{k}^{(j)} \big|a_{k}^{(j)}\rmv,\V{z}_{1:k} \big)$ in \eqref{eq:JPDA_approx3}, \eqref{eq:JPDA_approx3a} is equal for all $a_{k}^{(j)}\rmv =\rmv 0,\ldots,M_k$. Furthermore, $p(\V{a}_{k}|\V{z}_{1:k})$ is invariant to a permutation of the entries $a^{(j)}_{k}$ of $\V{a}_{k}$. This, in turn, implies that the marginal posterior \ac{da} \ac{pmf}s $p\big(a^{(j)}_{k} \big|\V{z}_{1:k}\big)$, $j \rmv=\rmv 1,\dots,L$ calculated from $p(\V{a}_{k}|\V{z}_{1:k})$ according to \eqref{eq:JPDA_marg2} are equal and, further, the approximate marginal state \acp{pdf} $\tilde{f}\big(\V{x}_{k}^{(j)} \big|\V{z}_{1:k}\big)$, $j \rmv= 1,\dots,L$ calculated from $p\big(a_{k}^{(j)} \big|\V{z}_{1:k}\big)$ and $f\big(\V{x}_{k}^{(j)} \big|a_{k}^{(j)}\rmv,\V{z}_{1:k} \big)$ according to \eqref{eq:JPDA_approx2} are equal as well. Therefore, the target state estimates $\V{x}^{(j) \text{MMSE}}_{k}$ calculated from $\tilde{f}\big(\V{x}_{k}^{(j)} \big|\V{z}_{1:k}\big)$ according to \eqref{eq:JPDA_MMSE} (with $f\big(\V{x}^{(j)}_k \big|\V{z}_{1:k} \big)$ replaced by $\tilde{f}\big(\V{x}_{k}^{(j)} \big|\V{z}_{1:k}\big)$) become equal, which means that the estimated tracks merge and, thus, the track coalescence effect is observed. An inspection of \eqref{eq:JPDA_approx2} and \eqref{eq:jointAssoc} shows that this indistinguishability of targets with the same prior information $f\big(\V{x}^{(j)}_{k} \big| \V{z}_{1:k-1}\big)$ is a direct consequence of the measurement model with measurement-origin uncertainty. We will illustrate this fact by considering two simple scenarios.

The first scenario demonstrates the track coalescence effect without using the Gaussian assumption underlying the JPDA filter. We consider a 1-D
state space and two targets that are close to each other. The target states are the targets' 1-D positions $x_k^{(j)} \in \{-1,1\}$ for $j \!=\! 1,2$. The targets are observed by a sensor that generates the measurements $z_k^{(1)} \!=\! 1$ and $z_k^{(2)} \!=\rmv -1$. Thus, we have $L_k \rmv=\rmv L \rmv =\rmv 2$ and $M_k\rmv =\rmv 2$. There are no clutter measurements and no measurement noise, and the detection probability is assumed to be one. We further assume that each measurement equals either one of the true target positions $x_k^{(j)}$ with equal probability, i.e., $z_{k}^{(1)} \!=\! 1$ equals $x_{k}^{(1)}$ or $x_{k}^{(2)}$ with equal probability $1/2$, and similarly for $z_{k}^{(2)} \!=\rmv -1$. \rd{From these assumptions, it follows that} $p\big(a_k^{(1)} = m \big| \V{z}_{1:k} \big) = p \big(a_k^{(2)} = m \big| \V{z}_{1:k}\big) = 1/2$ for $m = 1,2$ \rd{and $f\big(\V{x}_{k}^{(j)} \big|a_{k}^{(j)} \rmv=\rmv 1,\V{z}_{1:k} \big) =  \delta\big(x_{k}^{(j)} \!\rmv-\! 1\big)$ and $f\big(\V{x}_{k}^{(j)} \big|a_{k}^{(j)} \rmv=\rmv 2,\linebreak 
\V{z}_{1:k} \big) =  \delta\big(x_{k}^{(j)} \!\rmv+\! 1\big)$ for $j = 1,2$. Inserting these expression into \eqref{eq:JPDA_approx2} yields} $\tilde{f}\big(x_{k}^{(j)} \big|\V{z}_{1:k}\big) = \frac{1}{2}\ist \delta\big(x_{k}^{(j)} \!\rmv-\! 1\big) + \frac{1}{2}\ist \delta\big(x_{k}^{(j)} \!\rmv+\! 1\big)$ for $j \!=\! 1,2$. Finally, using \eqref{eq:JPDA_MMSE}, the MMSE state estimates are obtained as $\hat{x}_k^{(j)\text{MMSE}} \!=\rmv 0$ for $j \!=\! 1,2$. This means that the tracks are merged and, thus, the track coalescence effect is observed.

For the second scenario, we reconsider the simple tracking scenario previously considered in the context of MHT methods in Section \ref{sec:MHT-Tracking}. We recall this scenario for convenience. There are two targets, whose states include the targets' positions. Each target generates one measurement, which is the target's position plus Gaussian measurement noise, and there are no clutter measurements. Thus, we again have $L_k \rmv=\rmv L \rmv =\rmv 2$ and $M_k\rmv =\rmv 2$. The targets move on parallel tracks and in close proximity, in the sense that \rd{the distance between them} is significantly smaller than the standard deviation of the measurement noise. 
If the two targets move in this way for a sufficiently long time, it can be expected that the predicted posterior pdfs $f\big(\V{x}_{k}^{(1)} \big|\V{z}_{1:k-1}\big)$ and $f\big(\V{x}_{k}^{(2)} \big|\V{z}_{1:k-1}\big)$, i.e., the ``prior information,'' are approximately equal and that each of the two measurements is approximately equally likely to have originated from either one of the two targets. 
This assumption results in $f\big(\V{x}_{k}^{(1)} \big|a_{k}^{(1)} \!\rmv=\rmv m,\V{z}_{1:k}\big)$ and $f\big(\V{x}_{k}^{(2)} \big|a_{k}^{(2)} \!\rmv= m,\V{z}_{1:k}\big)$ in \eqref{eq:JPDA_approx3} being approximately equal for $m \rmv=\rmv 1,2$ and $p\big(a_{k}^{(1)} = m \big|\V{z}_{1:k} \big)$ and $p\big(a_{k}^{(2)} = m \big|\V{z}_{1:k} \big)$ being also approximately equal for $m \rmv=\rmv 1,2$.
As a consequence, it can be expected that,
according to \eqref{eq:JPDA_approx2}, $\tilde{f}\big(\V{x}_{k}^{(1)} \big|\V{z}_{1:k}\big)$ and $\tilde{f}\big(\V{x}_{k}^{(2)} \big|\V{z}_{1:k}\big)$ are almost equal.
As explained above, this implies that the target state estimates $\V{x}^{(1) \text{MMSE}}_{k}$ and $\V{x}^{(2) \text{MMSE}}_{k}$ become approximately equal, which means that the track coalescence effect is observed. The track coalescence effect also occurs when one of the two targets is missed, i.e., $M_k\! =\! 1$, or there is clutter, i.e., $M_k \!>\! 2$, as long as the distance between the targets is smaller than the standard deviation of the measurement noise. In both cases, again, each of the measurements is approximately equally likely to have originated from either one of the two targets. If the measurements follow a more complicated, possibly nonlinear or non-Gaussian measurement model, the track coalescence effect can still be observed, but a characterization of  
the case where the targets are ``in close proximity'' may be more difficult.

\begin{figure*}[t!]
\hspace{6mm}
{\scalebox{0.9}{\hspace{-12mm} \includegraphics[scale=.9]{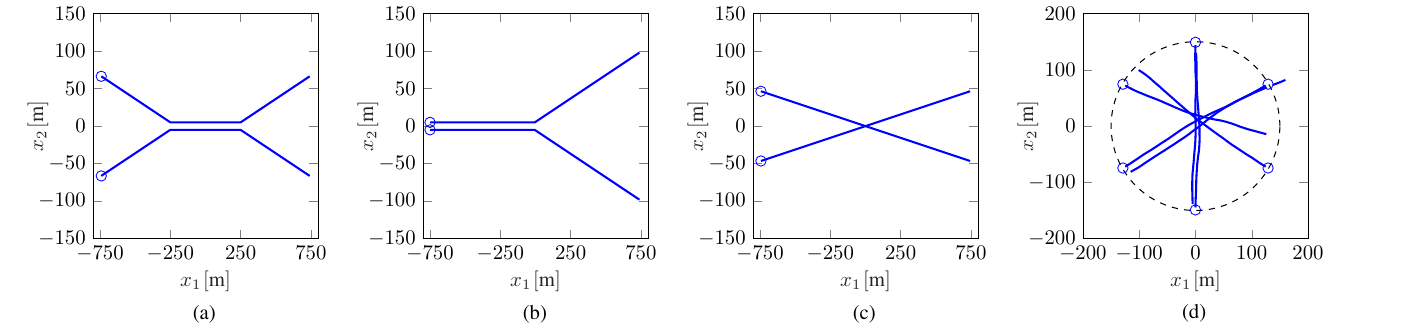}}}
\vspace{-5mm}
\caption{Target trajectories of scenarios (a) S1, (b) S2, (c) S3, and (d) S4. The initial target positions are indicated by circles.}
\label{fig:GroundTruth}
\vspace{-3mm}
\end{figure*}

\vspace{-1mm} 

\subsection{BP Method}
\label{sec:closerlook_bp}

Although the BP method considered in Section \ref{sec:BP-Tracking} is partly related to the JPDA filtering paradigm, it exhibits significantly reduced track coalescence effects. There are two reasons for this fact: (i) the BP method typically uses a particle representation of the marginal posterior pdfs $f\big(\V{x}^{(j)}_{k} \big| r^{(j)}_{k} \!\rmv=\! 1, \V{z}_{1:k} \big)$ in \eqref{eq:BP-post} \cite{mey17}, which preserves the multimodality of these pdfs, and (ii) the overconfident nature of the DA solution provided by the BP method favors the most likely PT-measurement association and thus a separation of target state estimates.

Reason (i) can be explained as follows. In the original JPDA filter, the \acp{pdf} $f\big(\V{x}_{k}^{(j)} \big|a_{k}^{(j)} \rmv,\V{z}_{1:k} \big)$ in \eqref{eq:JPDA_approx3} and \eqref{eq:JPDA_approx3a} are Gaussian pdfs. As a consequence, the approximate marginal posterior \acp{pdf} $\tilde{f}\big(\V{x}_{k}^{(j)} \big|\V{z}_{1:k}\big)$ calculated according to \eqref{eq:JPDA_approx2} are Gaussian mixture pdfs. The JPDA filter approximates these multimodal Gaussian mixture \acp{pdf} by Gaussian \acp{pdf}. This additional approximation exacerbates the track coalescence effect since \acp{pdf} $\tilde{f}\big(\V{x}_{k}^{(j)} \big|\V{z}_{1:k}\big)$ that are well distinguishable may become indistinguishable after the Gaussian approximation. By contrast, the particle representation of the marginal posterior pdfs $f\big(\V{x}^{(j)}_{k} \big| r^{(j)}_{k} \!\rmv=\! 1, \V{z}_{1:k} \big)$ that is used in the BP method is able to capture the multimodality of $f\big(\V{x}^{(j)}_{k} \big| r^{(j)}_{k} \!\rmv=\! 1, \V{z}_{1:k} \big)$. We note that the role of $f\big(\V{x}^{(j)}_{k} \big| r^{(j)}_{k} \!\rmv=\! 1, \V{z}_{1:k} \big)$ in the BP method is similar to that of $f\big(\V{x}^{(j)}_{k} \big| \V{z}_{1:k} \big)$ in JPDA filter methods.

Reason (ii) is the overconfident nature of the BP method for a factor graph with loops (``loopy BP'') \cite{WaiJor:B08}. To illustrate this aspect, 
we consider modified marginal posterior \ac{da} \acp{pmf} $p_{\hspace{-.15mm} \rho}\big(a^{(j)}_k \big|\V{z}_{1:k}\big) \triangleq \big( p\big(a^{(j)}_k \big|\V{z}_{1:k}\big)\big)^{\rho} \rmv/\ist C^{(j)}_k$ with \vspace{0.4mm} exponent $\rho \rmv\in\rmv (0, \infty)$, where $C^{(j)}_k \triangleq \sum^{M_k}_{a^{(j)}_k \!=\ist 0} \rmv \big( p\big(a^{(j)}_k \big|\V{z}_{1:k}\big) \big)^{\rho}\rmv$ \vspace{-0.9mm} and $p\big(a^{(j)}_k \big|\V{z}_{1:k}\big)$ is calculated according to \eqref{eq:JPDA_marg2}. For $0 \rmv<\rmv \rho \rmv<\rmv 1$, $p_{\hspace{-0.15mm} \rho}\big(a^{(j)}_k \big|\V{z}_{1:k}\big)$ is underconfident relative to $p\big(a^{(j)}_k \big|\V{z}_{1:k}\big)$, in the sense that likely PT-measurement associations (corresponding to large values of $p\big(a^{(j)}_k \big|\V{z}_{1:k}\big)$) are deemphasized and unlikely ones (corresponding to small values of $p\big(a^{(j)}_k \big|\V{z}_{1:k}\big)$) are emphasized, resulting in a larger spread of the modified \ac{pmf} $p_{\hspace{-.15mm} \rho}\big(a^{(j)}_k \big|\V{z}_{1:k}\big)$. In particular, for $\rho \rmv\rightarrow\rmv 0$, $p_{\hspace{-0.15mm} \rho}\big(a^{(j)}_k \big|\V{z}_{1:k}\big)$ is the uniform \ac{pmf}, i.e., all PT-measurement associations are equally likely. For $\rho \rmv>\rmv 1$, on the other hand, $p_{\hspace{-0.15mm} \rho}\big(a^{(j)}_k \big|\V{z}_{1:k}\big)$ is overconfident relative to $p\big(a^{(j)}_k \big|\V{z}_{1:k}\big)$, in the sense that likely associations are emphasized and unlikely ones are deemphasized, resulting in a smaller spread of $p_{\hspace{-.15mm} \rho}\big(a^{(j)}_k \big|\V{z}_{1:k}\big)$. In particular, for $\rho \rmv\rightarrow\rmv \infty$, $p_{\hspace{-0.15mm} \rho}\big(a^{(j)}_k \big|\V{z}_{1:k}\big)$ is one for the value of $a^{(j)}_k$ where $p\big(a^{(j)}_k \big|\V{z}_{1:k}\big)$ is largest---i.e., the MAP estimate of $\rv{a}^{(j)}_k$---and zero otherwise. The JPDA filter performs soft \ac{da} with $\rho \!=\! 1$, which leads to track coalescence as discussed in Section \ref{sec:closerlook_jpda}. By contrast, MHT methods perform hard \ac{da}, in that they use only the MAP estimate of $\rv{a}^{(j)}_k$ instead of the entire posterior distribution of $\rv{a}^{(j)}_k\rmv$, corresponding to $p_{\hspace{-0.15mm} \rho}\big(a^{(j)}_k \big|\V{z}_{1:k}\big)$ with $\rho \rmv\rightarrow\rmv \infty$; this leads to track repulsion as explained in Section \ref{sec:MHT-Tracking}.

The BP method, just as JPDA filtering methods, performs soft \ac{da} in the sense that it calculates approximate marginal posterior state \acp{pdf} $\tilde{f}\big(\V{x}_{k}^{(j)} \big|\V{z}_{1:k}\big)$ via a weighted summation over all possible associations, thereby taking into account the entire \ac{da} distribution (see \eqref{eq:JPDA_approx2}). However, in contrast to JPDA filter methods, it relies on loopy BP and thus computes overconfident approximations $\tilde{p}\big(a^{(j)}_k \big|\V{z}_{1:k}\big)$ of the marginal posterior \ac{da} \acp{pmf} $p\big(a^{(j)}_k \big|\V{z}_{1:k}\big)$. These approximations resemble $p_{\hspace{-0.15mm} \rho}\big(a^{(j)}_k \big|\V{z}_{1:k}\big)$ for some $\rho \!>\! 1$, which means that the BP method emphasizes likely PT-measurement associations and deemphasizes unlikely ones. Thus, the soft \ac{da} performed by the BP method is somewhat closer, in regard to the tracking results, to the hard \ac{da} performed by MHT methods (which do not exhibit the track coalescence effect). This explains why the BP method exhibits a reduced track coalescence effect. However, the BP solution is still less confident than the solution provided by MHT, which does not reflect association ambiguity by setting $\rho = \infty$. The price paid for the reduced track coalescence is that the BP solution tends to be overconfident.

\section{Simulation Study}
\label{sec:results}

Next, we present simulation results assessing and comparing the performance of the considered MTT methods in four different scenarios where targets come in close proximity. In particular, we will demonstrate experimentally that the BP method exhibits no track repulsion effect and a significantly reduced track coalescence effect compared to JPDA filter methods.

\begin{figure*}[t!]
\hspace{13mm}
	{\scalebox{0.9}{\hspace{-12mm} \includegraphics[scale=.8]{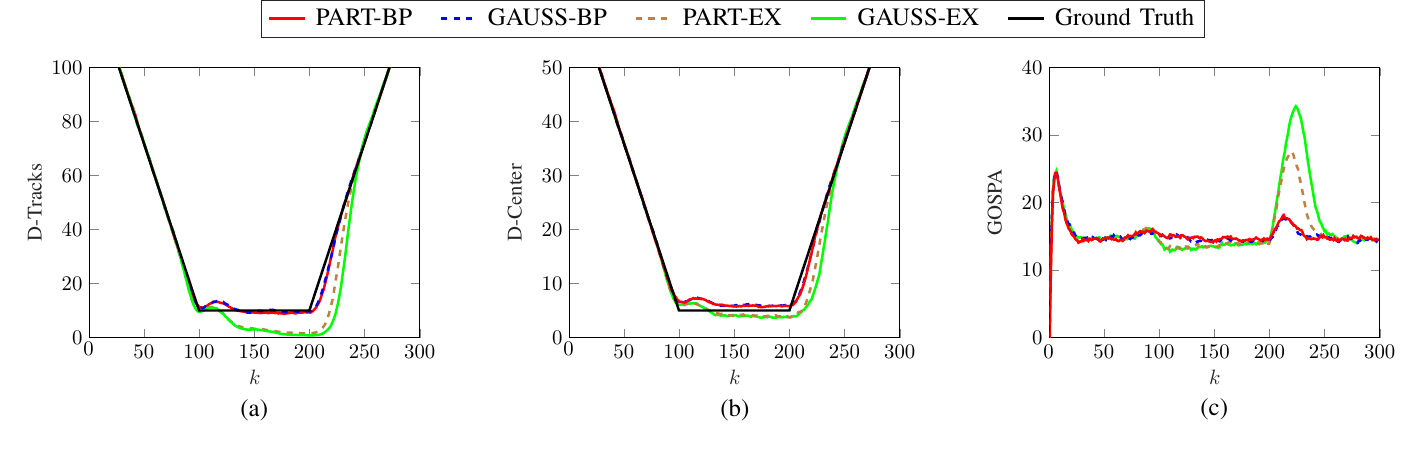}}}
\vspace{-2mm}
\caption{Performance of PART-BP, GAUSS-BP, PART-EX, and GAUSS-EX for scenario S1: 
(a) D-Tracks, (b) D-Center, and (c) GOSPA.}
\label{fig:BPvsEx}
\vspace{0mm}
\end{figure*}

\vspace{-1mm} 

\subsection{Simulation Setup}
\label{sec:results_setupt}

We consider the four simulation scenarios S1 through S4 depicted in Fig.\ \ref{fig:GroundTruth}. In S1 through S3, there are two targets with deterministically chosen trajectories: in S1, the two targets approach each other, move in parallel close to each other, and separate again; in S2, they move in parallel close to each other from the beginning and then separate; and in S3, they cross each other. In S4, on the other hand, six targets are born with uniform spacing on a circle of radius 150\ist\text{m} and evolve randomly according to a nearly constant-velocity motion (NCVM) model \cite{barShalom02}; they come in close proximity near the origin and then separate again \cite{mey17}. Each scenario comprises 300 time steps. The minimum distance between the two targets in S1 and S2 is equal to the standard deviation of the measurement noise, $\sigma_{\V{v}}$ (defined below). The targets remain close to each other for 100 time steps in S1 and for 150 time steps in S2. In S3 and S4, the two and six targets are in a ``2$\sigma_{\V{v}}$-neighborhood'' for about 50 time steps and 30 time steps, respectively. The region of interest (ROI) is $[-750\ist\text{m}, \ist 750\ist\text{m} ]  \times [-750\ist\text{m}, \ist 750\ist\text{m}]$ for all scenarios. 

The target states consist of 2-D position and velocity, i.e., $\RV{x}_{k}^{(j)} \rmv= \big[\rv{x}_{1,k}^{(j)} \;\ist \rv{x}_{2,k}^{(j)} \;\ist \dot{\rv{x}}_{1,k}^{(j)} \;\ist \dot{\rv{x}}_{2,k}^{(j)}\big]^{\text{T}}\rmv$. The various MTT methods, for all scenarios S1 through S4, use the NCVM model \cite{barShalom02}
\begin{equation}
\label{eq:MeasMod}
\RV{x}_{k}^{(j)} = \M{A}\ist\RV{x}^{(j)}_{k-1} + \RV{u}_{k}^{(j)} \rmv,
\end{equation}
where $\RV{u}_{k}^{(j)} \!\sim\! \Set{N}(\V{0},\M{\Sigma}_{\V{u}} )$  is a sequence of independent and identically distributed (iid) 4-D Gaussian random vectors and $\M{A}$ and $\M{\Sigma}_{\V{u}}$ are given by 
\[
\small
\M{A} = 
{ \begin{pmatrix}
  1 & 0 & T & 0 \\
   0 & 1 & 0 & T \\
   0  & 0  & 1 & 0 \\
   0 & 0 & 0 & 1
  \end{pmatrix} } \ist,
\quad\, \M{\Sigma}_{\V{u}} =
 { \begin{pmatrix}
   \frac{T^3}{3} & 0 & \frac{T^2}{2} & 0\\
   0 & \frac{T^3}{3} & 0 & \frac{T^2}{2}\\
   \frac{T^2}{2} & 0 & T & 0\\
   0 & \frac{T^2}{2} & 0 & T
  \end{pmatrix} } \ist \sigma^2_{\V{u}}\ist.
  \vspace{0mm}
\]
Here, $T\rmv=\rmv1\ist\text{s}$ and $\sigma^2_{\V{u}} \!=\rmv 0.1 \ist\ist \text{m}^2/\text{s}^4$ for S1, S2, and S3 and $\sigma^2_{\V{u}} \!= 0.0001 \ist\ist \text{m}^2/\text{s}^4$ for S4. \rd{We note that the changes in direction occurring in the target trajectories in S1 and S2 are not directly modeled by the NCVM model in \eqref{eq:MeasMod}; however, this model mismatch is compensated by the relatively high driving noise variance $\sigma^2_{\V{u}}$ used by the MTT methods in S1 and S2.} The survival probability is set to $p_{\text{s}} = 0.995$. Furthermore, $\mu_{\text{b}} = 0.01$ and $f_{\text{b}}(\overline{\V{x}}_{k})$ is uniform over the ROI.

The sensor produces noisy measurements of the target positions. More specifically, the target-generated measurements are given\vspace{-2mm} by
\[
\RV{z}_{k}^{(m)} =\ist   [\rv{x}_{1,k} \;\ist \rv{x}_{2,k} ]^{\text{T}} \rmv+\ist \RV{v}_{k}^{(m)} \rmv,
\vspace{0mm}
\]
where $\RV{v}_{k}^{(m)} \!\sim\! \Set{N}(\V{0},\sigma_{\V{v}}^2 {\bf I}_2)$ with $\sigma_{\V{v}} \!=\! 10\,$m is an iid sequence of 2-D Gaussian random vectors. The detection probability is set to $p_{\text{d}} \rmv=\rmv 0.5$. We assume a sensor with perfect resolution, i.e., independently of their distance, the targets can always be resolved in the sense that two different targets do not lead to the same measurement. In addition to the target-generated measurements, there are also clutter measurements. The mean number of clutter measurements is $\mu_{\text{c}} \!\rmv=\!\rmv 10$, and the clutter pdf $f_{\text{c}}\big( \V{z}_{k}^{(m)} \big)$ is uniform on the ROI. \rd{These measurement parameters and state evolution parameters are used in the various MTT methods; furthermore, the same measurement parameters and, in scenario S4, also the NCVM state evolution parameters are used to generate the measurements. We note that if larger values of $\sigma^2_{\V{u}}$, $\sigma^2_{\V{v}}$, and $\mu_{\text{c}}$ and smaller values of $p_{\text{s}}$ and $p_{\text{d}}$ are used to generate the measurements, then a poorer tracking accuracy of the various MTT methods can be expected.}

For each of our four scenarios, we performed 1000 simulation runs, each comprising 300 time steps. For performance evaluation, we use the \ac{gospa} metric \cite{RahGarSve:C17} based on the L2-norm, averaged over all simulation runs, with parameters $p\rmv=\rmv1$, $c\rmv=\rmv50$, and $\beta \rmv =\rmv 2$. The \ac{gospa} metric accounts for both cardinality and state estimation errors, similar to the OSPA metric \cite{schuhmacher08} and the complete OSPA metric \cite{Vu20COSPA}. For S1, we compute two additional performance metrics. The first, termed ``D-Tracks,'' is the distance between the two tracks, i.e., $|\hat{x}_{2,k}^{(1)} - \hat{x}_{2,k}^{(2)}|$, for each time step $k$. The second, termed ``D-Center,'' is the average of the distances of the estimated y-coordinates of the two tracks from the y-center (origin), i.e., $(|\hat{x}_{2,k}^{(1)}| + |\hat{x}_{2,k}^{(2)}|)/2$, again for each time step $k$. A small D-Tracks metric can indicate track coalescence, while the D-Center metric assesses the accuracy of the tracks' centroid (which is high if the D-Center metric is small). For example, if D-Tracks and D-Center are both small, the estimated tracks may have coalesced, but at least the estimated ``joint track'' is quite accurate. On the other hand, if D-Center is large, then also the estimated joint track is inaccurate.

\subsection{Reduced Track Coalescence in the BP Method} 
\label{sec:results_BP}

We first demonstrate and analyze experimentally the reduction of track coalescence exhibited by the BP method of Section \ref{sec:BP-Tracking}. In Section \ref{sec:closerlook_bp}, we argued that this reduction is due to the particle representation of the marginal posterior state \acp{pdf}---which, differently from a Gaussian representation, preserves multimodality---and the overconfident nature of the DA-related messages.
We now investigate this numerically by combining two choices of track representation (single Gaussian and set of particles) with the DA strategies employed by the BP method and the JPDA filter. In particular, to demonstrate the influence of the particle representation of the marginal posterior state \acp{pdf}, we compare a particle implementation of the BP method with a Gaussian implementation; these implementations will be designated as PART-BP and GAUSS-BP, respectively. Furthermore, to demonstrate the influence of the overconfident nature of the DA-related messages in the BP method,
we consider modified versions of the particle and Gaussian implementations in which the BP-based approximate marginalization of the association pmf is replaced by an exact marginalization. These latter methods will be designated as PART-EX and GAUSS-EX. To exclude the influence of target detection errors, we temporarily assume that the birth times and birth positions of the targets are perfectly known by the various tracking algorithms. We note that in this case GAUSS-EX coincides with the JPDA filter of Section \ref{sec:JPDA-Tracking}, \rd{and thus the corresponding implementation equations of GAUSS-EX can be found, e.g., in \cite{barShalom11}. Furthermore, we recall that in GAUSS-BP, the exact marginalization of GAUSS-EX is replaced by the BP-based approximate marginalization.} PART-BP and PART-EX use 5000 particles to represent the PT states. \rd{The implementation equations of PART-BP are equal to those in \cite{mey17} for the case of a single sensor and known birth times and locations. We recall that in PART-EX, the BP-based marginalization of PART-BP is replaced by the exact marginalization.} The threshold for target confirmation is $P_{\text{th}} \rmv=\rmv 0.5$. 

Fig.~\ref{fig:BPvsEx} presents the results for scenario S1. The D-Tracks curves in Fig.~\ref{fig:BPvsEx}(a) show that the tracks of GAUSS-BP and PART-BP are close to the ground truth, whereas those of GAUSS-EX and PART-EX tend to merge and exhibit a delay in separating again. This behavior is also reflected by the GOSPA curves in Fig.~\ref{fig:BPvsEx}(c): the track coalescence effect causes an increase of the GOSPA error of GAUSS-EX and PART-EX for $k \rmv=\rmv 200,\ldots,250$. This increase is smaller for PART-EX than for GAUSS-EX. Regarding the D-Center curves in Fig.~\ref{fig:BPvsEx}(b), all four methods exhibit a similar behavior. In summary, the results in Figs.~\ref{fig:BPvsEx}(a) and (c) demonstrate that both 
particle representation and BP-based marginalization tend to lead to 
reduced track coalescence effects, thereby confirming our reasoning of Section \ref{sec:closerlook}.

\begin{figure}[t!]
\hspace{7mm}
{\scalebox{0.75}{\hspace{-13.5mm} \includegraphics[scale=.8]{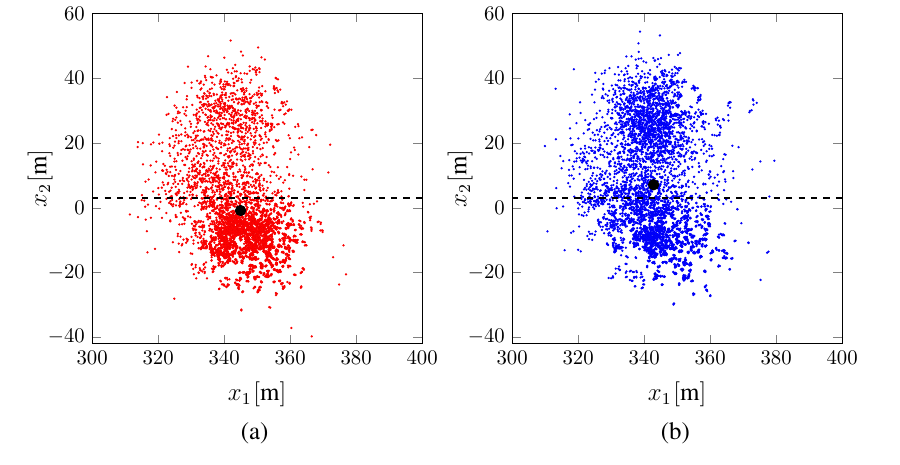}}}
\vspace{-5.5mm}
\caption{\rd{Example realizations of particle sets representing the marginal posterior pdfs (a) $f(\V{x}_k^{(1)}|\V{z}_{1:k})$
and (b) $f(\V{x}_k^{(2)}|\V{z}_{1:k})$ in PART-EX at time $k = 220$, for scenario S1. The black dots indicate the corresponding
state estimates $\hat{\V{x}}_{220}^{(1)}$ and $\hat{\V{x}}_{220}^{(2)}$. The dashed straight line corresponds to
$x_2 = (\hat{x}_{2,220}^{(1)} + \hat{x}_{2,220}^{(2)})/2$\ist.}}
\label{fig:PART_EX}
\vspace{-2mm}
\end{figure}

\rd{Finally, we demonstrate that the particle representation preserves multimodality of the marginal posterior pdfs. For scenario S1, we plot in Fig.\ \ref{fig:PART_EX} example realizations of the two particle sets employed by PART-EX to represent the marginal posterior pdfs $f(\V{x}_k^{(1)}|\V{z}_{1:k})$ and $f(\V{x}_k^{(2)}|\V{z}_{1:k})$ at time $k = 220$, as well as the corresponding state estimates $\hat{\V{x}}_{220}^{(1)}$ and $\hat{\V{x}}_{220}^{(2)}$. It can be seen that the particle sets express a bimodality of the marginal posterior pdfs. 
Furthermore, a comparison of the spatial separation between the state estimates $\hat{\V{x}}_{220}^{(1)}$ and $\hat{\V{x}}_{220}^{(2)}$ with the true separation of the targets (which is $10\text{m}$) suggests that the track coalescence effect is not significant. We note that for a reduction of track coalescence, the particle representation of the marginal posterior pdfs is especially beneficial in PART-EX, whereas in PART-BP, the overconfident nature of the BP messages is predominant.}

\begin{figure*}[t!]
\hspace{13mm}
{\scalebox{0.9}{\hspace{-12mm} \includegraphics[scale=.8]{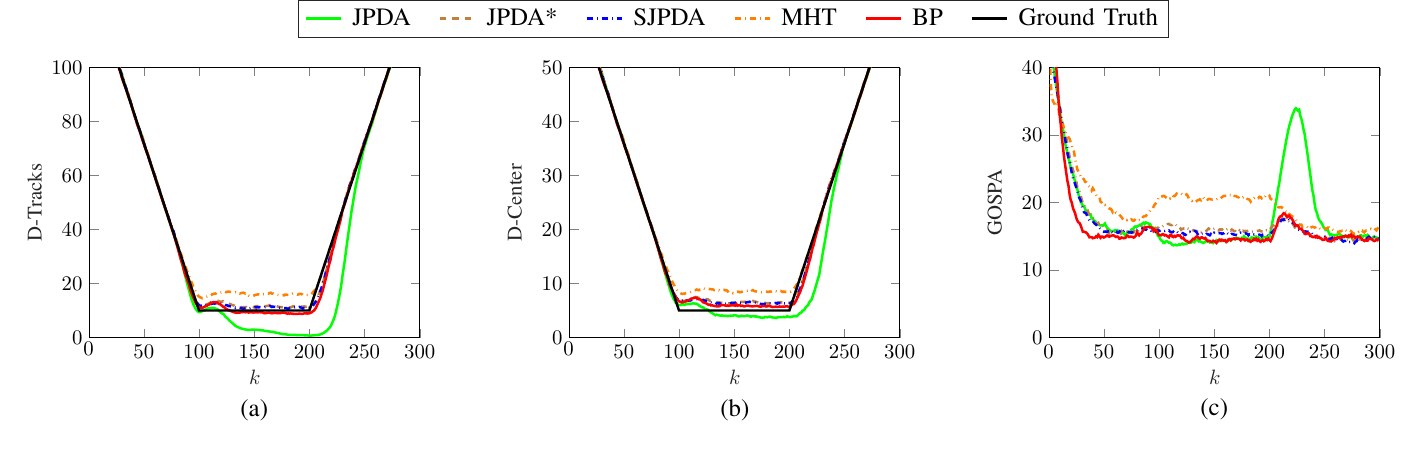}}}
\vspace{-2mm}
\caption{Performance of JPDA, JPDA*, SJPDA, MHT, and BP for scenario S1: (a) D-Tracks, (b) D-Center, and (c) GOSPA.}
\label{fig:BPvsMHTvsJPDA}
\vspace{-1mm}
\end{figure*}

\vspace{-1mm}

\subsection{Analysis of Track Coalescence and Repulsion for the\\JPDA, JPDA*, SJPDA, MHT, and BP Methods}
\label{sec:results_an}

Next, we analyze the track coalescence and repulsion effects potentially exhibited by the JPDA, JPDA*, SJPDA, (track-oriented) MHT, and BP methods for our four scenarios S1 through S4. The number of targets and the target birth times and positions are now unknown to all methods. JPDA, JPDA*, and SJPDA use gating with a gate validation threshold of 13.82, corresponding to an in-gate probability (i.e., the probability that a target-generated measurement is within the gate) of 0.999. We remark that a gate validation threshold of 9.21, corresponding to an in-gate probability of 0.99, led to similar results.
MHT is a reference implementation provided by \textit{Systems \& Technology Research}, \rd{Woburn, MA, USA}; this implementation uses a Gaussian representation of the individual tracks \cite{CroWilBar:C11} and 
a hypothesis depth of five scans. 
The track confirmation logics \cite{barShalom11} of JPDA, JPDA*, and SJPDA are set to 12/24, those of MHT to 8/16 (these values were chosen such that each method achieves its best performance across all scenarios). Finally, BP generates a new track for each measurement, sets the corresponding existence probability to $10^{-4}\rmv$, and prunes existing tracks with existence probabilities below $10^{-4}\rmv$. 

Fig.~\ref{fig:BPvsMHTvsJPDA} shows the D-Tracks, D-Center, and GOSPA curves for scenario S1. One can see in Fig.~\ref{fig:BPvsMHTvsJPDA}(a) that when the two targets move in close proximity, D-Tracks is increased for MHT and decreased for JPDA compared to the ground truth tracks; this indicates track repulsion and track coalescence, respectively. The low D-Tracks curve of JPDA for $k \rmv=\rmv 200,\ldots,240$ demonstrates that, after they coalesced, the tracks estimated by JPDA separate again only with a delay. In JPDA*, SJPDA, and BP, track coalescence is strongly reduced. In the case of JPDA* and SJPDA, this is due to the special design of these methods; in the case of BP, it is due to the particle representation of the marginal posterior state pdfs and the overconfident nature of the DA-related messages in BP 
as argued in Section \ref{sec:closerlook_bp} and verified experimentally in Section \ref{sec:results_BP}. 
We note that JPDA*, as BP, performs approximate averaging of the DA vectors, which leads to overconfident marginal posterior DA pmfs \cite{bloblo00}.

Furthermore, in Fig.~\ref{fig:BPvsMHTvsJPDA}(b), the D-Center curves of JPDA, JPDA*, SJPDA, and BP are generally close to the true curves. An exception is JPDA for $k \rmv=\rmv 200,\ldots,240$, which again reflects the fact that, after coalescence, the tracks estimated by JPDA separate again only with a delay. The GOSPA curves shown in Fig.~\ref{fig:BPvsMHTvsJPDA}(c) confirm the results of Figs.~\ref{fig:BPvsMHTvsJPDA}(a) and (b). In particular, the increased GOSPA error of MHT for $k \rmv=\rmv 100,\ldots,200$ is due to track repulsion, and that of JPDA for $k \rmv=\rmv 200,\ldots,240$ to track coalescence. JPDA*, SJPDA, and BP perform similarly well, with a slight performance advantage for BP.

\begin{figure}[t!]
\hspace{7mm}
\vspace{2mm}
{\scalebox{0.65}{\hspace{-16mm} \includegraphics[scale=1]{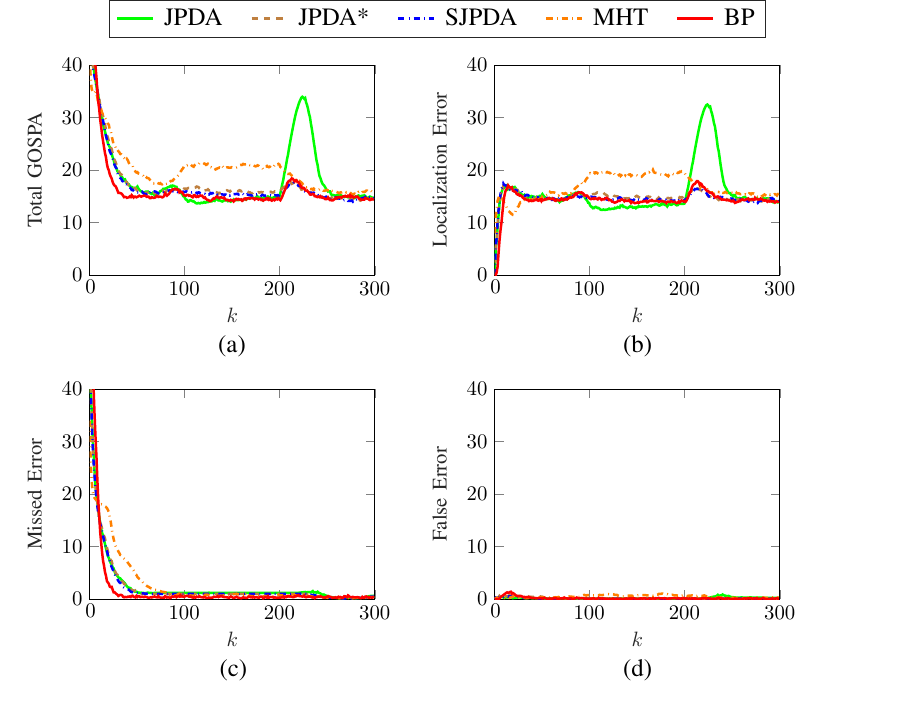}}}
\vspace{-9mm}
\caption{GOSPA error decomposition for scenario S1: (a) Total GOSPA error (same as Fig.~\ref{fig:BPvsMHTvsJPDA}(c)),
(b) localization error component, (c) missed error component, and (d) false error component.}
\label{fig:AllGospaS1}
\vspace{-2mm}
\end{figure}

\begin{figure}[t!]
\hspace{7mm}
\vspace{2mm}
{\scalebox{0.65}{\hspace{-16mm} \includegraphics[scale=1]{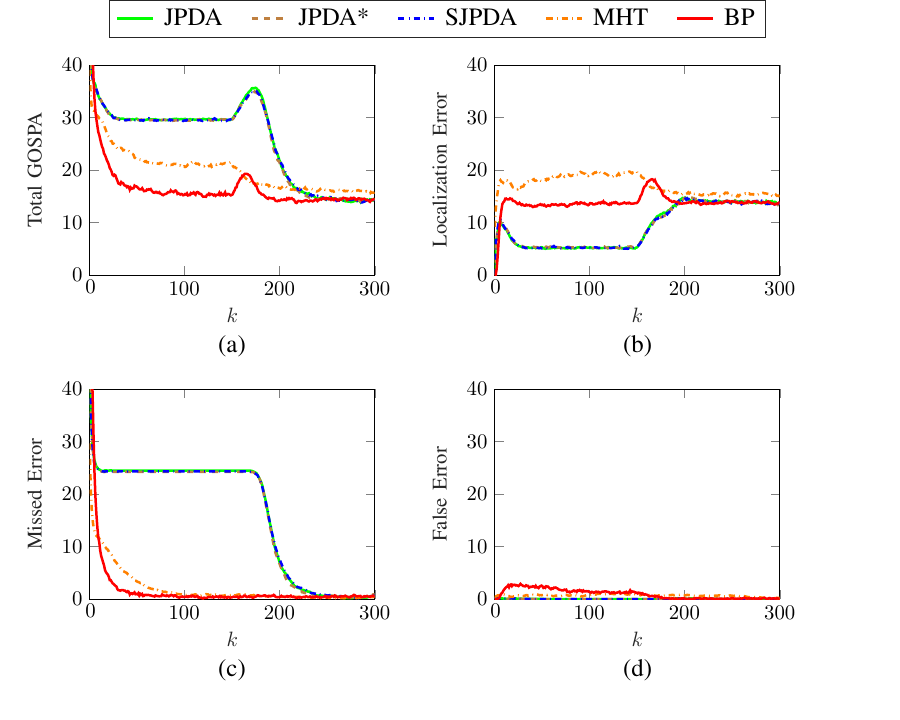}}}
\vspace{-9mm}
\caption{Total GOSPA error (a) and its components (b)--(d) for scenario S2.}
\label{fig:AllGospaS2}
\vspace{-3mm}
\end{figure}

Still for scenario S1, Figs.~\ref{fig:AllGospaS1}(b)--(d) show the individual GOSPA error components, i.e., the localization error, the error due to missed targets (``missed error''), and the error due to false targets (``false error'') \cite{RahGarSve:C17}. In addition, the total GOSPA curves from Fig.~\ref{fig:BPvsMHTvsJPDA}(c) are replicated in Fig.~\ref{fig:AllGospaS1}(a) for easy reference. The localization error curves in Fig.~\ref{fig:AllGospaS1}(b) are seen to be similar to the total GOSPA curves; this reflects the fact that missed and false targets contribute only little to the total GOSPA error.

Fig.~\ref{fig:AllGospaS2} shows the GOSPA error and its components for scenario S2. In S2, differently from S1, the two targets are in close proximity right from the beginning (see Fig.~\ref{fig:GroundTruth}(b)). This poses a challenge for the track initiation stage of the JPDA filter methods. Indeed, JPDA, JPDA*, and SJPDA generate only one track as long as the targets remain in close proximity. This means that one of the targets is missed, resulting in a large missed error component of JPDA, JPDA*, and SJPDA in the time range $k \rmv=\rmv 1,\ldots,200,$ as shown in Fig.~\ref{fig:AllGospaS2}(c). By contrast, as also shown in Fig.~\ref{fig:AllGospaS2}(c), MHT and BP initialize both tracks correctly. Fig.~\ref{fig:AllGospaS2}(b) indicates an increased localization error of MHT for $k \rmv=\rmv 1,\ldots,150$; this is due to track repulsion as in S1. It is furthermore seen that the localization error of JPDA, JPDA*, and SJPDA is reduced; this can be explained by the fact that these methods estimate just a single track whereas the other methods estimate both tracks, combined with the fact that GOSPA is an ``unnormalized'' error metric, which implies that tracking fewer targets generally results in a lower localization error. Note, however, that the total GOSPA error is lowest for BP. Finally, Fig.~\ref{fig:AllGospaS2}(d) shows that the false error component is small for all methods.

\begin{figure}[t!]
\hspace{7mm}
{\scalebox{0.65}{\hspace{-16mm} \includegraphics[scale=1]{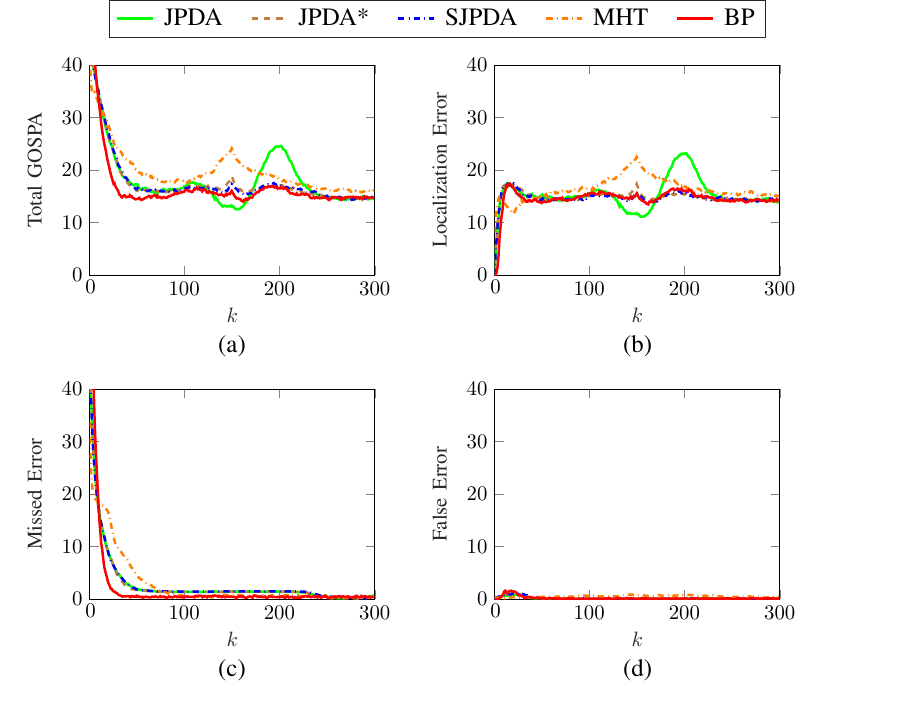}}}
\vspace{-6.5mm}
\caption{Total GOSPA error (a) and its components (b)--(d) for scenario S3.}
\label{fig:AllGospaS3}
\vspace{-1mm}
\end{figure}

Fig.~\ref{fig:AllGospaS3} presents the GOSPA error and its components for scenario S3. Similarly to scenarios S1 and S2, it can be seen that when the two targets are close to each other, i.e., for $k \rmv=\rmv 125,\ldots,175$, MHT suffers from track repulsion, and after the targets separate again, i.e., for $k \rmv=\rmv 175,\ldots,225$, JPDA suffers from track coalescence. By contrast, track repulsion and track coalescence are nonexistent or significantly reduced in JPDA*, SJPDA, and BP.

\begin{figure}[t!]
\hspace{7mm}
\vspace{2mm}
{\scalebox{0.65}{\hspace{-16mm} \includegraphics[scale=1]{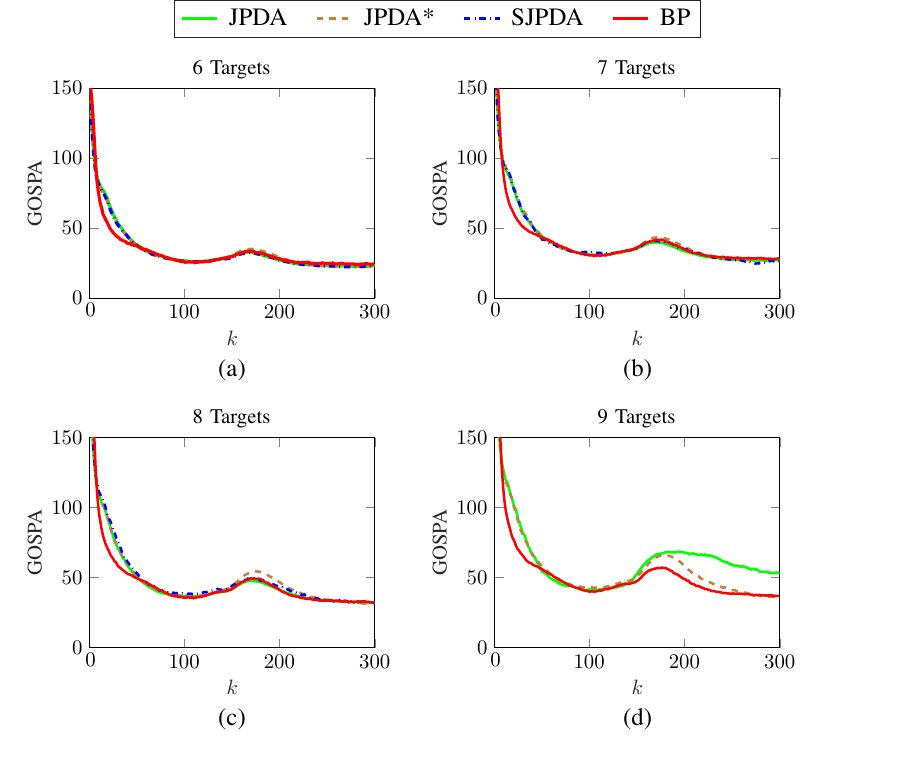}}}
\vspace{-8mm}
\caption{GOSPA error for scenario S4 (a) and for three variants of S4 with seven, eight, and nine targets (b)--(d).}
\label{fig:AllGospaS4}
\vspace{-2.5mm}
\end{figure}

\subsection{Further Analysis of Track Coalescence for the\\JPDA, JPDA*, SJPDA, and BP Methods}
\label{sec:results_S4}
\vspace{1mm}

The results reported so far for scenarios S1 through S3 demonstrate that JPDA, JPDA*, SJPDA, and BP do not exhibit track repulsion and MHT does not exhibit track coalescence. In addition, JPDA*, SJPDA, and BP exhibit reduced track coalescence compared to conventional JPDA. Since JPDA*, SJPDA, and BP performed equally well, we now further analyze the track coalescence behavior of these methods and, for comparison, of conventional JPDA in the even more challenging scenario S4. Since MHT is not susceptible to track coalescence, it is not considered in this analysis. S4 features six targets, which come in close proximity around time $k \!=\! 150$ (see Fig.~\ref{fig:GroundTruth}(d)). We also consider more challenging variants of S4 with seven, eight, and nine targets. The GOSPA error for these scenarios is shown in Fig.~\ref{fig:AllGospaS4}. It can be seen that all four methods exhibit only small track coalescence effects for the cases of six, seven, and eight targets. For the case of nine targets, JPDA and JPDA* exhibit an increased GOSPA error compared to BP. In particular, the GOSPA error of JPDA remains high for a long time after the targets separate again.

\begin{table}
\vspace{1.5mm}
\centering
\vspace*{0mm}
{\small
\hspace*{-0.5mm}	\begin{tabular}{|l|l|rrrr|}	
	\hline
  \rule{0mm}{3mm}  
	& & \hspace{-1mm}6 Targets\hspace{-1mm} & 7 Targets\hspace{-1mm} & 8 Targets\hspace{-1mm} & 9 Targets \\[0mm] \hline
	\rule{0mm}{3mm}\multirow{4}{*}{\hspace{-1mm}JPDA\hspace{-1mm}}   
	& \hspace{-1mm}min    & 0.54\,s\hspace{-1mm}  &  0.90\,s\hspace{-1mm}  &    2.05\,s\hspace{-1mm}  &   18.76\,s  \\[.2mm]
	& \hspace{-1mm}median & 1.42\,s\hspace{-1mm}  &  4.54\,s\hspace{-1mm}  &   17.45\,s\hspace{-1mm}  &   84.14\,s  \\[.2mm]
	& \hspace{-1mm}mean   & 1.44\,s\hspace{-1mm}  &  4.59\,s\hspace{-1mm}  &   33.28\,s\hspace{-1mm}  & 2952.90\,s  \\[.2mm]
	& \hspace{-1mm}max    & 3.43\,s\hspace{-1mm}  & 18.34\,s\hspace{-1mm}  & 2758.00\,s\hspace{-1mm}  & 9458.90\,s  \\[0mm]\hline 
	\rule{0mm}{3mm}\multirow{4}{*}{\hspace{-1mm}JPDA*\hspace{-1mm}}       
	& \hspace{-1mm}min    & 0.52\,s\hspace{-1mm}  &  0.90\,s\hspace{-1mm}  &   2.42\,s\hspace{-1mm}  &   16.31\,s  \\[.2mm]
	& \hspace{-1mm}median & 1.22\,s\hspace{-1mm}  &  4.89\,s\hspace{-1mm}  &  34.66\,s\hspace{-1mm}  &  256.08\,s  \\[.2mm]
	& \hspace{-1mm}mean   & 1.29\,s\hspace{-1mm}  &  5.35\,s\hspace{-1mm}  &  39.70\,s\hspace{-1mm}  &  309.18\,s  \\[.2mm]
	& \hspace{-1mm}max    & 3.62\,s\hspace{-1mm}  & 22.26\,s\hspace{-1mm}  & 340.30\,s\hspace{-1mm}  & 1653.30\,s  \\[0mm]\hline  
	\rule{0mm}{3mm}\multirow{4}{*}{\hspace{-1mm}SJPDA\hspace{-1mm}}      
	& \hspace{-1mm}min   &  1.28\,s\hspace{-1mm}  &  16.21\,s\hspace{-1mm}  &  132.57\,s\hspace{-1mm}  &       \\[.2mm]
	& \hspace{-1mm}median& 10.30\,s\hspace{-1mm}  &  70.63\,s\hspace{-1mm}  &  531.93\,s\hspace{-1mm}  &       \\[.2mm]
	& \hspace{-1mm}mean  & 10.46\,s\hspace{-1mm}  &  74.90\,s\hspace{-1mm}  & 640.120\,s\hspace{-1mm}  &       \\[.2mm]
	& \hspace{-1mm}max   & 36.73\,s\hspace{-1mm}  & 251.72\,s\hspace{-1mm}  & 4351.40\,s\hspace{-1mm}  &       \\[0mm]\hline 	
	\rule{0mm}{3.2mm}\multirow{4}{*}{\hspace{-1mm}BP\hspace{-1mm}} 
	& \hspace{-1mm}min    & 18.62\,s\hspace{-1mm}  & 19.33\,s\hspace{-1mm}  & 21.03\,s\hspace{-1mm}  & 21.76\,s \\[.2mm]	
	& \hspace{-1mm}median & 24.04\,s\hspace{-1mm}  & 24.84\,s\hspace{-1mm}  & 27.78\,s\hspace{-1mm}  & 31.18\,s \\[.2mm]
	& \hspace{-1mm}mean   & 24.05\,s\hspace{-1mm}  & 24.86\,s\hspace{-1mm}  & 27.86\,s\hspace{-1mm}  & 31.26\,s \\[.2mm]
	& \hspace{-1mm}max    & 28.44\,s\hspace{-1mm}  & 29.88\,s\hspace{-1mm}  & 31.80\,s\hspace{-1mm}  & 36.48\,s \\[0mm]\hline	
	\end{tabular}
}
\vspace{.5mm}
\caption{Minimum, median, mean, and maximum runtimes of JPDA, JPDA*, SJPDA, and BP for scenario S4 (six targets) and its variants with seven, eight, and nine targets.}
\label{fig:runtimes} 
\vspace{-3mm}
\end{table}

Still considering scenario S4 and its variants, we report in Table~\ref{fig:runtimes} the minimum, median, mean, and maximum total (for all 300 time steps) filter runtimes of Matlab implementations of JPDA, JPDA*, SJPDA, and BP on an Intel Xeon Gold 5222 CPU averaged over 1000 simulation runs. Our JPDA, JPDA*, SJPDA, and BP implementations use gating and clustering to reduce computational complexity. However, in Scenario S4 and its variants, this has only a limited effect during the time the targets are in close proximity because the targets are within the same gate and thus clustered together. As Table~\ref{fig:runtimes} shows, the runtimes of JPDA, JPDA*, and SJPDA increase rapidly with the number of targets. This is consistent with the exponential scaling of the complexity of these filters in the number of tracks; note that the number of tracks may be higher than the number of targets because of false tracks generated due to clutter measurements. Because of the exponential scaling, JPDA and JDPA* are infeasible for more than nine targets and SJPDA is infeasible for more than eight targets. Table~\ref{fig:runtimes} also shows that for JPDA, JPDA*, and SJPDA, there is a large difference between the minimum and maximum filter runtimes. The high filter runtimes in large scenarios are due to simulation runs where the number of tracks is considerably higher than the actual number of targets. Furthermore, because there are significantly  more simulation runs with a low filter runtime than with a high filter runtime, the median runtimes differ significantly from the mean runtimes. On the other hand, the runtime of BP increases only slowly with the number of targets. This is consistent with the fact that the complexity of BP scales only linearly in the number of targets. \vspace{-1mm}

\section{Conclusion}
\label{sec:conclusion}
\vspace{0mm}

We reviewed and analyzed three major methodologies for multitarget tracking: the classical joint probabilistic data association (JPDA) filter and multiple hypothesis tracking (MHT) methodologies and the recently introduced belief propagation (BP) methodology. The focus of our study was on track coalescence and track repulsion effects, which are well known to compromise the performance of, respectively, JPDA filter and MHT methods when targets are in close proximity. In particular, we investigated the potential occurrence of track coalescence and track repulsion effects in the BP method. We argued that the BP method does not suffer from track repulsion because it performs soft data association similarly to the JPDA filter. Moreover, track coalescence effects in the BP method are significantly smaller than in the JPDA filter, because certain properties of the BP messages related to data association encourage separation of target state estimates. These theoretical arguments were confirmed by simulation experiments. Our numerical results demonstrated excellent performance of the BP method compared to the JPDA, JPDA*, SJPDA, and MHT methods in scenarios with targets in close proximity. 
Future work will include an analysis of the quality of the BP state estimates in terms of the normalized estimation error squared (NEES) \cite{Bar98}.

\renewcommand{\baselinestretch}{.946}
\footnotesize
\selectfont
\bibliographystyle{IEEEtran}
\bibliography{references,IEEEabrv,StringDefinitions}

\begin{thebibliography}{10}
\providecommand{\url}[1]{#1}
\csname url@samestyle\endcsname
\providecommand{\newblock}{\relax}
\providecommand{\bibinfo}[2]{#2}
\providecommand{\BIBentrySTDinterwordspacing}{\spaceskip=0pt\relax}
\providecommand{\BIBentryALTinterwordstretchfactor}{4}
\providecommand{\BIBentryALTinterwordspacing}{\spaceskip=\fontdimen2\font plus
\BIBentryALTinterwordstretchfactor\fontdimen3\font minus
  \fontdimen4\font\relax}
\providecommand{\BIBforeignlanguage}[2]{{%
\expandafter\ifx\csname l@#1\endcsname\relax
\typeout{** WARNING: IEEEtran.bst: No hyphenation pattern has been}%
\typeout{** loaded for the language `#1'. Using the pattern for}%
\typeout{** the default language instead.}%
\else
\language=\csname l@#1\endcsname
\fi
#2}}
\providecommand{\BIBdecl}{\relax}
\BIBdecl

\bibitem{barShalom11}
Y.~Bar-Shalom, P.~K. Willett, and X.~Tian, \emph{{Tracking and Data Fusion: A
  Handbook of Algorithms}}.\hskip 1em plus 0.5em minus 0.4em\relax Storrs, CT,
  USA: Yaakov Bar-Shalom, 2011.

\bibitem{reid79}
D.~B. Reid, ``An algorithm for tracking multiple targets,'' \emph{IEEE Trans.
  Autom. Control}, vol.~24, no.~6, pp. 843--854, Dec. 1979.

\bibitem{ChaMor11}
S.~Challa, M.~R. Morelande, D.~Mu{\v s}icki, and R.~J. Evans,
  \emph{{Fundamentals of Object Tracking}}.\hskip 1em plus 0.5em minus
  0.4em\relax Cambridge, UK: Cambridge University Press, 2011.

\bibitem{mahler2007statistical}
R.~Mahler, \emph{{Statistical Multisource-Multitarget Information
  Fusion}}.\hskip 1em plus 0.5em minus 0.4em\relax Norwood, MA, USA: Artech
  House, 2007.

\bibitem{koch14}
W.~Koch, \emph{{Tracking and Sensor Data Fusion: Methodological Framework and
  Selected Applications}}.\hskip 1em plus 0.5em minus 0.4em\relax Berlin,
  Germany: Springer, 2014.

\bibitem{mey18}
F.~Meyer, T.~Kropfreiter, J.~L. Williams, R.~Lau, F.~Hlawatsch, P.~Braca, and
  M.~Z. Win, ``Message passing algorithms for scalable multitarget tracking,''
  \emph{Proc. IEEE}, vol. 106, no.~2, pp. 221--259, Feb. 2018.

\bibitem{barShalom74}
Y.~Bar-Shalom, ``Extension of the probabilistic data association filter in
  multi-target tracking,'' in \emph{Proc. SNETA-74}, San Diego, CA, USA, Sep.
  1974.

\bibitem{bar-shalom09}
Y.~Bar-Shalom, F.~Daum, and J.~Huang, ``The probabilistic data association
  filter,'' \emph{IEEE Control Syst. Mag.}, vol.~29, pp. 82--100, Dec. 2009.

\bibitem{fortmann83}
T.~Fortmann, Y.~Bar-Shalom, and M.~Scheffe, ``Sonar tracking of multiple
  targets using joint probabilistic data association,'' \emph{IEEE J. Ocean.
  Eng.}, vol.~8, no.~3, pp. 173--184, Jul. 1983.

\bibitem{FerMunTesBraMeyPelPetAlvStrLeP:J17}
G.~Ferri, A.~Munaf\`{o}, A.~Tesei, P.~Braca, F.~Meyer, K.~Pelekanakis,
  R.~Petroccia, J.~Alves, C.~Strode, and K.~LePage, ``Cooperative robotic
  networks for underwater surveillance: {A}n overview,'' \emph{IET Radar Sonar
  Navig.}, vol.~11, no.~12, pp. 1740--1761, Dec. 2017.

\bibitem{Kim21}
D.~Y. {Kim}, B.~{Ristic}, R.~{Guan}, and L.~{Rosenberg}, ``A {B}ernoulli
  track-before-detect filter for interacting targets in maritime radar,''
  \emph{IEEE Trans. Aerosp. Electron. Syst.}, vol.~57, no.~3, pp. 1--10, Jan.
  2021.

\bibitem{Gag20}
D.~Gaglione, G.~Soldi, F.~Meyer, F.~Hlawatsch, P.~Braca, A.~Farina, and M.~Z.
  Win, ``Bayesian information fusion and multitarget tracking for maritime
  situational awareness,'' \emph{IET Radar, Sonar Navig,}, vol.~14, no.~12, pp.
  1845--1857, Oct. 2020.

\bibitem{levinson11}
J.~Levinson \emph{et~al.}, ``Towards fully autonomous driving: Systems and
  algorithms,'' in \emph{Proc. IEEE IV 2011}, Baden-Baden, Germany, Jun. 2011,
  pp. 163--168.

\bibitem{urmson08}
C.~Urmson \emph{et~al.}, ``Autonomous driving in urban environments: Boss and
  the urban challenge,'' \emph{J. Field Robot., Special Issue on the 2007 DARPA
  Urban Challenge, Part I}, vol.~25, no.~8, pp. 425--466, Jun. 2008.

\bibitem{MeyWil:J21}
F.~Meyer and J.~L. Williams, ``Scalable detection and tracking of geometric
  extended objects,'' \emph{IEEE Trans. Signal Process.}, vol.~69, pp.
  6283--6298, 2021.

\bibitem{Bar:B90}
Y.~Bar-Shalom, \emph{{Multitarget-Multisensor Tracking: Advanced
  Applications}}.\hskip 1em plus 0.5em minus 0.4em\relax Norwood, MA, USA:
  Artech House, 1990.

\bibitem{musicki04}
D.~Musicki and R.~Evans, ``{Joint integrated probabilistic data association:
  JIPDA},'' \emph{IEEE Trans. Aerosp. Electron. Syst.}, vol.~40, no.~3, pp.
  1093--1099, Jul. 2004.

\bibitem{VerMas05}
J.~Vermaak, S.~Maskell, and M.~Briers, ``A unifying framework for multi-target
  tracking and existence,'' in \emph{Proc. FUSION-05}, Philadelphia, PA, USA,
  Jul. 2005, pp. 250--258.

\bibitem{MusLaS08}
D.~Musicki and B.~La~Scala, ``Multi-target tracking in clutter without
  measurement assignment,'' \emph{IEEE Trans. Aerosp. Electron. Syst.},
  vol.~44, no.~3, pp. 877--896, Jul. 2008.

\bibitem{horridge06}
P.~Horridge and S.~Maskell, ``{Real-time tracking of hundreds of targets with
  efficient exact JPDAF implementation},'' in \emph{Proc. FUSION-06}, Florence,
  Italy, Jul. 2006.

\bibitem{PatPop00}
K.~Pattipati, R.~L. Popp, and T.~Kirubarajan, ``Survey of assignment techniques
  for multitarget tracking,'' in \emph{{Multitarget-Multisensor Tracking:
  Applications and Advances}}, Y.~Bar-Shalom and W.~D. Blair, Eds.\hskip 1em
  plus 0.5em minus 0.4em\relax Norwood, MA, USA: Artech House, 2000, vol.~3,
  ch.~2, pp. 77--159.

\bibitem{Cox96}
I.~J. Cox and S.~L. Hingorani, ``{An efficient implementation of Reid's
  multiple hypothesis tracking algorithm and its evaluation for the purpose of
  visual tracking},'' \emph{IEEE Trans. Pattern Anal. Mach. Intell.}, vol.~18,
  no.~2, pp. 138--150, Feb. 1996.

\bibitem{DanNew06}
R.~Danchick and G.~E. Newnam, ``{Reformulating Reid's MHT method with
  generalised Murty K-best ranked linear assignment algorithm},'' \emph{IEE
  Radar Sonar Navig.}, vol. 153, no.~1, pp. 13--22, Feb. 2006.

\bibitem{Bla04}
S.~S. Blackman, ``Multiple hypothesis tracking for multiple target tracking,''
  \emph{IEEE Trans. Aerosp. Electron. Syst.}, vol.~19, no.~1, pp. 5--18, Jan.
  2004.

\bibitem{Mor77}
C.~Morefield, ``Application of 0-1 integer programming to multitarget tracking
  problems,'' \emph{IEEE Trans. Autom. Control}, vol.~22, no.~3, pp. 302--312,
  Jun 1977.

\bibitem{deb97}
S.~Deb, M.~Yeddanapudi, K.~Pattipati, and Y.~Bar-Shalom, ``{A generalized S-D
  assignment algorithm for multisensor-multitarget state estimation},''
  \emph{IEEE Trans. Aerosp. Electron. Syst.}, vol.~33, no.~2, pp. 523--538,
  Apr. 1997.

\bibitem{ZhaMey:J24}
W.~Zhang and F.~Meyer, ``Multisensor multiobject tracking with improved
  sampling efficiency,'' \emph{IEEE Trans. Signal Process.}, vol.~72, pp.
  2036--2053, 2024.

\bibitem{LiaMey:J24}
M.~Liang and F.~Meyer, ``Neural enhanced belief propagation for multiobject
  tracking,'' \emph{IEEE Trans. Signal Process.}, vol.~72, pp. 15--30, Sep.
  2024.

\bibitem{Mah03}
R.~P.~S. Mahler, ``{Multitarget Bayes filtering via first-order multitarget
  moments},'' \emph{IEEE Trans. Aerosp. Electron. Syst.}, vol.~39, no.~4, pp.
  1152--1178, Oct. 2003.

\bibitem{vo05}
B.-N. Vo, S.~Singh, and A.~Doucet, ``{Sequential Monte Carlo methods for
  multitarget filtering with random finite sets},'' \emph{IEEE Trans. Aerosp.
  Electron. Syst.}, vol.~41, no.~4, pp. 1224--1245, Oct. 2005.

\bibitem{vo07}
B.-T. Vo, B.-N. Vo, and A.~Cantoni, ``Analytic implementations of the
  cardinalized probability hypothesis density filter,'' \emph{IEEE Trans.
  Signal Process.}, vol.~55, no.~7, pp. 3553--3567, Jul. 2007.

\bibitem{williams2015marg}
J.~L. Williams, ``{Marginal multi-Bernoulli filters: RFS derivation of MHT,
  JIPDA and association-based MeMBer},'' \emph{IEEE Trans. Aerosp. Electron.
  Syst.}, vol.~51, no.~3, pp. 1664--1687, Jul. 2015.

\bibitem{Nannuru16}
S.~Nannuru, S.~Blouin, M.~Coates, and M.~Rabbat, ``{Multisensor CPHD filter},''
  \emph{IEEE Trans. Aerosp. Electron. Syst.}, vol.~52, no.~4, pp. 1834--1854,
  Aug. 2016.

\bibitem{Gar20SoT}
{\'A}.~F. Garc{\'i}a-Fern{\'a}ndez, L.~{Svensson}, and M.~R. {Morelande},
  ``Multiple target tracking based on sets of trajectories,'' \emph{IEEE Trans.
  Aerosp. Electron. Syst.}, vol.~56, no.~3, pp. 1685--1707, Jun. 2020.

\bibitem{Gra18SoTPMBM}
K.~{Granstr\"om}, L.~{Svensson}, Y.~{Xia}, J.~{Williams}, and {\'A}.~F.
  Garc{\'i}a-Fern{\'a}ndez, ``{P}oisson multi-{B}ernoulli mixture trackers:
  {C}ontinuity through random finite sets of trajectories,'' in \emph{Proc.
  FUSION-18}, Cambridge, UK, Jul. 2018, pp. 973--981.

\bibitem{KroMeyHla:J20}
T.~Kropfreiter, F.~Meyer, and F.~Hlawatsch, ``A fast labeled multi-{B}ernoulli
  filter using belief propagation,'' \emph{IEEE Trans. Aerosp. Electron.
  Syst.}, vol.~56, no.~3, pp. 2478--2488, Jun. 2020.

\bibitem{KroMeyHla:J22}
------, ``An efficient labeled/unlabeled random finite set algorithm for
  multiobject tracking,'' \emph{IEEE Trans. Aerosp. Electron. Syst.}, vol.~58,
  no.~6, pp. 5256--5275, Dec. 2022.

\bibitem{Ber:B91}
D.~P. Bertsekas, \emph{Linear Network Optimization: Algorithms and
  Codes}.\hskip 1em plus 0.5em minus 0.4em\relax Cambridge, MA, USA: {MIT}
  Press, 1991.

\bibitem{Sal:J90}
D.~Salmond, ``Mixture reduction algorithms for target tracking in clutter,'' in
  \emph{Proc. SPIE-90}, Orlando, FL, USA, Apr. 1990, pp. 434--445.

\bibitem{Pao:94}
L.~Y. Pao, ``Multisensor multitarget mixture reduction algorithms for
  tracking,'' \emph{J. Guid. Control Dyn.}, vol.~17, no.~6, pp. 1205--1211,
  Nov. 1994.

\bibitem{vermaak05}
J.~Vermaak, S.~J. Godsill, and P.~Perez, ``{Monte Carlo filtering for multi
  target tracking and data association},'' \emph{IEEE Trans. Aerosp. Electron.
  Syst.}, vol.~41, no.~1, pp. 309--332, Jan. 2005.

\bibitem{mey17}
F.~Meyer, P.~Braca, P.~Willett, and F.~Hlawatsch, ``A scalable algorithm for
  tracking an unknown number of targets using multiple sensors,'' \emph{IEEE
  Trans. Signal Process.}, vol.~65, no.~13, pp. 3478--3493, Mar. 2017.

\bibitem{fitzgerald85}
R.~J. Fitzgerald, ``Track biases and coalescence with probabilistic data
  association,'' \emph{IEEE Trans. Signal Process.}, vol.~21, no.~6, pp.
  822--825, Nov. 1985.

\bibitem{corcar09}
S.~Coraluppi, C.~Carthel, P.~Willett, M.~Dingboe, O.~O'Neill, and T.~Luginbuhl,
  ``The track repulsion effect in automatic tracking,'' in \emph{Proc.
  FUSION-09}, Seattle, WA, USA, Jul. 2009, pp. 2225--2230.

\bibitem{bloblo00}
H.~A.~P. Blom and E.~A. Bloem, ``Probabilistic data association avoiding track
  coalescence,'' \emph{IEEE Trans. Autom. Control}, vol.~45, no.~2, pp.
  247--259, Feb. 2000.

\bibitem{bloblo15}
H.~A.~P. Blom, E.~A. Bloem, and D.~Musicki, ``{JIPDA}: {Automatic} target
  tracking avoiding track coalescence,'' \emph{IEEE Trans. Aerosp. Electron.
  Syst.}, vol.~51, no.~2, pp. 962--974, Apr. 2015.

\bibitem{corcar12}
S.~Coraluppi and C.~Carthel, ``An equivalence-class approach to
  multiple-hypothesis tracking,'' in \emph{Proc. AeroConf-12}, Big Sky, MT,
  USA, Mar. 2012, pp. 1--8.

\bibitem{SveSve11}
L.~Svensson, D.~Svensson, M.~Guerriero, and P.~Willett, ``{Set JPDA filter for
  multitarget tracking},'' \emph{IEEE Trans. Signal Process.}, vol.~59, no.~10,
  pp. 4677--4691, Oct. 2011.

\bibitem{Wil14}
J.~L. Williams, ``An efficient, variational approximation of the best fitting
  multi-{B}ernoulli filter,'' \emph{IEEE Trans. Signal Process.}, vol.~63,
  no.~1, pp. 258--273, Jan. 2015.

\bibitem{LauWil16}
R.~A. Lau and J.~L. Williams, ``A structured mean field approach for
  existence-based multiple target tracking,'' in \emph{Proc. FUSION-16},
  Heidelberg, Germany, Jul. 2016, pp. 1111--1118.

\bibitem{reuter14}
S.~Reuter, B.-T. Vo, B.-N. Vo, and K.~Dietmayer, ``{The labeled multi-Bernoulli
  filter},'' \emph{IEEE Trans. Signal Process.}, vol.~62, no.~12, pp.
  3246--3260, Jun. 2014.

\bibitem{CroWilBar:C11}
S.~Coraluppi and C.~Carthel, ``Track management in multiple-hypothesis
  tracking,'' in \emph{Proc. IEEE SAM-18}, Sheffield, UK, Jul. 2018, pp.
  3840--3843.

\bibitem{CroWilGue:C11}
D.~F. Crouse, P.~Willett, M.~Guerriero, and L.~Svensson, ``{An approximate
  minimum MOSPA estimator},'' in \emph{Proc. IEEE ICASSP-11}, Prague, Czech
  Republic, Jul. 2011, pp. 3644--3647.

\bibitem{williams14}
J.~L. Williams and R.~Lau, ``Approximate evaluation of marginal association
  probabilities with belief propagation,'' \emph{IEEE Trans. Aerosp. Electron.
  Syst.}, vol.~50, no.~4, pp. 2942--2959, Oct. 2014.

\bibitem{SolMeyBraHla:J19}
G.~Soldi, F.~Meyer, P.~Braca, and F.~Hlawatsch, ``Self-tuning algorithms for
  multisensor-multitarget tracking using belief propagation,'' \emph{IEEE
  Trans. Signal Process.}, vol.~67, no.~15, pp. 3922--3937, Aug. 2019.

\bibitem{kschischang01}
F.~R. Kschischang, B.~J. Frey, and H.-A. Loeliger, ``{Factor graphs and the
  sum-product algorithm},'' \emph{IEEE Trans. Inf. Theory}, vol.~47, no.~2, pp.
  498--519, Feb. 2001.

\bibitem{koller09}
D.~Koller and N.~Friedman, \emph{{Probabilistic Graphical Models: Principles
  and Techniques}}.\hskip 1em plus 0.5em minus 0.4em\relax Cambridge, MA, USA:
  MIT Press, 2009.

\bibitem{Kro21BP}
T.~Kropfreiter, F.~Meyer, S.~Coraluppi, C.~Carthel, R.~Mendrzik, and
  P.~Willett, ``Track coalescence and repulsion: {MHT}, {JPDA}, and {BP},'' in
  \emph{Proc. FUSION-21}, Sun City, South Africa, Nov. 2021.

\bibitem{MorCho:C16}
S.~Mori, C.-Y. Chong, and K.~C. Chang, ``Evaluation of data association
  hypotheses: {Non-Poisson} i.i.d. cases,'' in \emph{Proc. FUSION-04},
  Stockholm, Sweden, Jun. 2004.

\bibitem{PatDeb92}
K.~R. Pattipati, S.~Deb, Y.~Bar-Shalom, and R.~B.~J. Washburn, ``A new
  relaxation algorithm and passive sensor data association,'' \emph{IEEE Trans.
  Autom. Control}, vol.~37, no.~2, pp. 198--213, Feb. 1992.

\bibitem{PooRij93}
A.~P. Poore and N.~Rijavec, ``{A Lagrangian relaxation algorithm for
  multidimensional assignment problems arising from multitarget tracking},''
  \emph{SIAM J. Optim.}, vol.~3, no.~3, pp. 544--563, 1993.

\bibitem{PooGad06}
A.~B. Poore and S.~Gadaleta, ``Some assignment problems arising from multiple
  target tracking,'' \emph{Math. Comput. Model.}, vol.~43, no. 9--10, pp.
  1074--1091, 2006.

\bibitem{CorGriDet:C06}
S.~Coraluppi, D.~Grimmett, and P.~de~Theije, ``Benchmark evaluation of
  multistatic trackers,'' in \emph{Proc. FUSION-06}, Florence, Italy, Jul.
  2006.

\bibitem{Willet07}
P.~Willett, T.~Luginbuhl, and E.~Giannopoulos, ``{MHT} tracking for crossing
  sonar targets,'' in \emph{Proc. SPIE}, San Diego, CA, USA, Sep. 2007, pp.
  469--480.

\bibitem{schuhmacher08}
D.~Schuhmacher, B.-T. Vo, and B.-N. Vo, ``A consistent metric for performance
  evaluation of multi-object filters,'' \emph{IEEE Trans. Signal Process.},
  vol.~56, no.~8, pp. 3447--3457, Aug. 2008.

\bibitem{Cro13}
D.~F. Crouse, ``{Advances in displaying uncertain estimates of multiple
  targets},'' in \emph{Proc. SPIE-13}, Baltimore, MD, USA, May 2013.

\bibitem{Cro11}
D.~F. Crouse, P.~Willett, and Y.~Bar-Shalom, ``Developing a real-time track
  display that operators do not hate,'' \emph{IEEE Trans. Signal Process.},
  vol.~59, no.~7, pp. 3441--3447, Apr. 2011.

\bibitem{Fit90}
R.~Fitzgerald, ``{Development of practical PDA logic for multitarget tracking
  by microprocessor},'' in \emph{{Multitarget-Multisensor Tracking: Advanced
  Applications}}, Y.~Bar-Shalom, Ed.\hskip 1em plus 0.5em minus 0.4em\relax
  Norwood, MA, USA: Artech-House, 1990.

\bibitem{MeyGem:J21}
F.~Meyer and K.~L. Gemba, ``Probabilistic focalization for shallow water
  localization,'' \emph{J. Acoust. Soc. Am.}, vol. 150, no.~2, pp. 1057--1066,
  Aug. 2021.

\bibitem{Gag21Fus}
D.~Gaglione, P.~Braca, G.~Soldi, F.~Meyer, F.~Hlawatsch, and M.~Z. Win,
  ``Fusion of sensor measurements and target-provided information in
  multitarget tracking,'' \emph{IEEE Trans. Signal Process.}, vol.~70, pp.
  322--336, Dec. 2021.

\bibitem{WeiFre:01}
Y.~Weiss and W.~T. Freeman, ``{Correctness of belief propagation in Gaussian
  graphical models of arbitrary topology},'' \emph{Neural Comput.}, vol.~13,
  no.~10, pp. 2173--2200, 2001.

\bibitem{WaiJor:B08}
M.~Wainwright and M.~Jordan, ``{Graphical models, exponential families, and
  variational inference},'' \emph{Foundations and Trends in Machine Learning},
  vol.~1, no. 1-2, pp. 1--305, 2008.

\bibitem{barShalom02}
Y.~Bar-Shalom, T.~Kirubarajan, and X.-R. Li, \emph{{Estimation with
  Applications to Tracking and Navigation}}.\hskip 1em plus 0.5em minus
  0.4em\relax New York, NY, USA: Wiley, 2002.

\bibitem{RahGarSve:C17}
A.~S. {Rahmathullah}, {\'A}.~F. Garc{\'i}a-Fern{\'a}ndez, and L.~{Svensson},
  ``Generalized optimal sub-pattern assignment metric,'' in \emph{Proc.
  FUSION-17}, Xi'an, China, Jul. 2017.

\bibitem{Vu20COSPA}
T.~Vu, ``A complete optimal subpattern assignment ({COSPA}) metric,'' in
  \emph{Proc. FUSION-20}, Rustenburg, South Africa, Jul. 2020.

\bibitem{Bar98}
Y.~Bar-Shalom and X.-R. Li, \emph{{Estimation and Tracking: Principles,
  Techniques, and Software}}.\hskip 1em plus 0.5em minus 0.4em\relax Norwood,
  MA, USA: Artech House, 1998.

\end{thebibliography}

\end{document}